\documentclass[pra,twocolumn,aps,floatfix,longbibliography,superscriptaddress]{revtex4-1}
\usepackage{graphicx}
\usepackage{amsmath}
\usepackage{amssymb}
\usepackage{amsfonts}
\usepackage{enumerate}
\usepackage{color}
\usepackage{tikz}
\usepackage{ragged2e}
\usetikzlibrary{calc,decorations.markings}
\usepackage[colorlinks=true,linkcolor=blue,citecolor=blue,urlcolor=blue]{hyperref}
\usepackage{tipa}
\usepackage{physics}
\usepackage{soul}
\usepackage{float}
\usepackage{amsthm}
\newtheorem*{nogotheorem}{No-go theorem}

\newcommand{\kk}{\boldsymbol{k}}
\newcommand{\xx}{\mathbf{x}}
\newcommand{\x}{\mathsf{x}}

\usepackage{tikz,xcolor}
\definecolor{lime}{HTML}{A6CE39}
\DeclareRobustCommand{\orcidicon}{%
	\begin{tikzpicture}
	\draw[lime, fill=lime] (0,0) 
	circle [radius=0.16] 
	node[white] {{\fontfamily{qag}\selectfont \tiny ID}};
	\draw[white, fill=white] (-0.0625,0.095) 
	circle [radius=0.007];
	\end{tikzpicture}
	\hspace{-2mm}
}
\foreach \x in {A, ..., Z}{%
	\expandafter\xdef\csname orcid\x\endcsname{\noexpand\href{https://orcid.org/\csname orcidauthor\x\endcsname}{\noexpand\orcidicon}}
}

\begin{document}

\title{An exactly solvable relativistic quantum Otto engine}
\author{Nikos K. Kollas\orcidA{}}
\email{kollas@upatras.gr}
\author{Dimitris Moustos\orcidB{}}
\email{dmoustos@upatras.gr}
\affiliation{Department of Physics, University of Patras, 26504 Patras, Greece}
\date{\today}

\begin{abstract}
We revisit the mathematics of exactly solvable Unruh-DeWitt detector models, interacting with massless scalar fields under instantaneous interactions, to construct a relativistic quantum Otto heat engine. By deriving the conditions under which the thermodynamic cycle is closed we study the effects of motion on the amount of work that can be extracted from the machine when the working medium is moving at a constant relativistic velocity through the heat baths. While there is a degrading effect with respect to speed in the hot bath, we demonstrate that in the case of the cold bath, genuine enhancing effects are sometimes present. For couplings the same order as the inverse frequency of the detector and a specific value for the temporal separation between the two instantaneous interactions--needed in order to be possible to cool the detector--a non-monotonic dependence between speed and extracted work exists raising the intriguing possibility of exploiting relativistic effects for the enhancement of thermodynamic processes in tabletop experiments. 
\end{abstract}
\maketitle
\section{Introduction}

Quantum thermodynamics \cite{goold2016role,vinjanampathy2016quantum,Lostaglio,binder2019thermodynamics,Deffner:Campell} investigates fundamental concepts, such as temperature, heat, and work in the quantum regime. A central focus of the field is the study of thermal machines \cite{Kosloff,Deffner:TMs,TMs:and:batt,cangemi2023quantum} designed to operate in the quantum realm, and whether quantum features can be harnessed in order to enhance their performance. 

Since the introduction of a three-level maser \cite{Scovil} as a prototype for a quantum heat engine, a number of theoretical investigations have emerged, exploring, among others, the impact of coherence \cite{Scully:Coh, QTM:coh}, squeezed or non-equilibrium thermal baths \cite{QTM:squeez,QTM:squeez:2,QTM:squeez:3}, non-Markovian effects \cite{Non:Mark:QTM,Non:Mark:QTM:2} and the strong coupling regime \cite{QTM:strong,QTM:strong:2,QTM:strong:3} on engine efficiency, with several proposals for their experimental realization \cite{exp:eng,Exp:eng:single:atom,Exp:engine:ion:spin,exp:spin:eng,Exp:QFT:machine:Huber} (for a more detailed overview
see \cite{Deffner:TMs} and references therein).

The implications of relativistic effects on the performance of quantum thermal machines has remained a relatively unexplored topic. Relativity, quantum physics and thermodynamics are known to be related through the Unruh effect \cite{Unruh, Takagi, Louko:Fewster,BJA:DM}, which asserts that a uniformly accelerated observer perceives the Minkowski vacuum of a quantum field as a thermal state at a temperature proportional to their acceleration. Motivated by this, the notion of a \emph{Unruh quantum Otto heat engine} has recently been introduced \cite{Arias_2018,Mann2018,Unruh:Otto:degen,Unruh:Otto:entangl,Unruh:Otto:entangl2,Unruh:Otto:reflect}.
Other studies have also explored the impact of relativistic energies and the effects of spacetime geometry on the thermal efficiency of thermodynamic cycles \cite{Otto:relativ:energ,Deffner:spacetime}. Moreover, quantum heat engines with confined relativistic fields as the working medium have been put forth as possible implementations with cavity-optomechanical setups \cite{DEB1,DEB2}.

In \cite{papadatos} the effects of relative motion of a hot reservoir with respect to the working medium in a quantum Otto engine were investigated, where it was reported that the amount of extracted work decreases with respect to the velocity of the bath. Leveraging the effective temperatures recorded by observers that move along stationary trajectories \cite{Effective:Unruh}, a general relativistic quantum Otto engine was introduced in \cite{Yoshimura:mann}, where the amount of work extracted by a circularly moving observer has been explicitly obtained.

A relativistic quantum Otto engine is treated within the framework of time-dependent perturbation theory, where an Unruh-DeWitt (UDW) detector moving along a trajectory in a background spacetime is weakly coupled to a scalar field \cite{Unruh,DeWitt,Hu:Louko}. Instead, by employing \emph{instantaneous} detector-field interactions it is possible to obtain an exact expression for the final state of the detector \cite{delta1,delta2,delta3,delta4,delta5,delta6,NK:DM:MM}. This allows for a complete investigation of the detector dynamics with respect to the full parameter space, such as the size and frequency of the detector and the strength of the coupling.

In the present work, we make use of instantaneous interactions to derive the necessary and sufficient conditions to close a thermodynamic Otto cycle. By considering a detector which is moving at a relativistic constant speed through the two baths, we observe that degradation effects still persist in the case of the hot bath \cite{papadatos}. However, for a detector with a size of the same order as its transition wavelength we find that, given a sufficient amount of time during which the detector cools, motion in the cold bath can positively enhance the performance of the engine. 

Throughout, we employ the natural system of units in which $\hbar=c=k_B=1$. We also assume a Minkowski spacetime with metric signature $(-+++)$. Four-vectors are represented by sans-serif characters ($\mathsf{x}$), while boldface letters ($\xx$) denote spatial vectors.

\section{Relativistic thermal engines with instantaneous interactions}

In a relativistic thermal engine the working medium consists of a two level \emph{Unruh-DeWitt detector} in a diagonal state of its energy basis
\begin{equation}\label{in-state}
	\rho = \frac{1-r}{2}|\Omega\rangle\langle\Omega|+\frac{1+r}{2}|0\rangle\langle 0|,
\end{equation}
with $0\leq r\leq 1$ its \emph{purity} \cite{PhysRevA.67.062104} and energy gap $\Omega$. The reservoirs are modeled as quantum fields at thermal states, which in the simplest case considered here consist of massless scalar fields
\begin{equation}
    \hat\varphi(\mathsf{x})=\int\frac{d^3\kk}{\sqrt{(2\pi)^32\abs{\kk}}}\left(\hat{a}_{\kk}e^{i\mathsf{k}\cdot\mathsf{x}}+\hat{a}^{\dagger}_{\kk}e^{-i\mathsf{k}\cdot\mathsf{x}}\right)
\end{equation}
with Hamiltonian
\begin{equation}
    \hat{H}_\phi=\int \abs{\kk}\hat a^\dagger_{\kk}\hat a_{\kk}d^3\kk,
\end{equation}
where $\hat a_{\kk}$ and $\hat a_{\kk}^\dag$ are the annihilation and creation operators of field mode $\kk$ satisfying the following canonical commutation relations
\begin{equation}\label{commut:rel}
    [\hat{a}_{\kk},\hat{a}_{\kk'}]=[\hat{a}_{\kk}^\dag,\hat{a}_{\kk'}^\dag]=0,\quad
    [\hat{a}_{\kk},\hat{a}_{\kk'}^\dag]=\delta(\kk-\kk').
\end{equation}

\begin{figure}
\centering
\includegraphics[width=0.6\columnwidth]{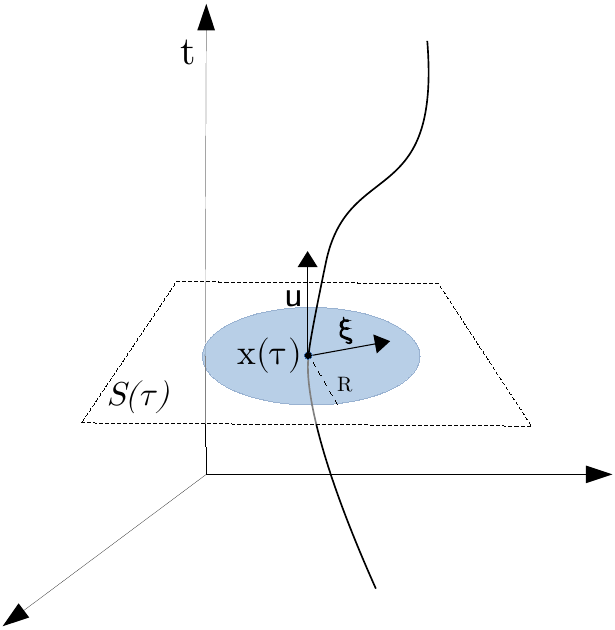}
\caption{An extended UDW detector interacts with the field not at a single point in spacetime but rather in a neighborhood of its position with an effective size $R$ given by the mean distance from its center of mass, weighted by an appropriate smearing function $f(\boldsymbol\xi)$, as defined in the detector's rest frame.}
\label{fig_smeared}
\end{figure}

The detector exchanges heat with the field by interacting with an \emph{instantaneous} UDW interaction of the form
\begin{equation}\label{smeared_interaction}
    \hat{H}_{\text{int}}(\tau)=\lambda\delta(t-t_0)\hat{\mu}(\tau)\otimes\hat{\varphi}_f(\mathsf{x}(\tau)),
\end{equation}
where $\lambda$ is a coupling constant with dimensions of length, $\hat{\mu}(\tau)$ is the \emph{transition operator} of the detector
\begin{equation}\label{monopole}
    \hat{\mu}(\tau)=e^{i\Omega \tau}|\Omega\rangle\langle 0|+e^{-i\Omega \tau}|0\rangle\langle\Omega|,
\end{equation}
and $\hat{\varphi}_f(\mathsf{x}(\tau))$ is a \emph{smeared field} along the detector's trajectory $\mathsf{x}(\tau)=(t(\tau),\xx(\tau))$ parameterized by its proper time $\tau$
\begin{equation}
\hat{\varphi}_f(\mathsf{x}(\tau))=\int_{\mathcal{S}(\tau)}f(\boldsymbol{\xi})\hat{\varphi}(\mathsf{x}(\tau)+\boldsymbol{\xi})d^3\boldsymbol\xi.
\end{equation}
Introducing a real valued \emph{smearing function} $f(\boldsymbol{\xi})$ makes it possible to take into account the spatial extension of the detector by averaging the field over a 3-dimensional timelike simultaneity hypersurfase $\mathcal{S}(\tau)$ at the detector's position (FIG. \ref{fig_smeared}) with an effective radius given by the weighted average distance
\begin{equation}
R=\int_{\mathcal S(\tau)}\abs{\boldsymbol\xi}f(\boldsymbol\xi)d^3\boldsymbol\xi.
\end{equation}
From a physical point of view the smearing function  reflects the shape and size of the detector \cite{Schlicht_2004,Louko_2006}, which may result from a point-like interaction of a hydrogen-like atom with the field when expanded in the detector's energy eigenfunctions \cite{smear:mart1,T:Rick}.

Suppose that at time $t_0$ a clock in the rest frame of the detector registers time equal to $\tau_0$, then for a field in a thermal state $\sigma_\varphi\propto e^{-\beta\hat{H}_\varphi}$ at inverse temperature $\beta$, the state of the detector after the interaction is given by the action of a bit flip channel 
\begin{equation}\label{final state}
    B(\rho)=(1-p)\rho+p\hat\mu(\tau_0)\rho\hat\mu(\tau_0)
\end{equation}
with
\begin{equation}\label{z}
    p=\frac{1}{2}(1-e^{-2\lambda^2\dot\tau^2_0\langle\hat\varphi^2_{f_0}\rangle_\beta}),
\end{equation}
where $\dot\tau_0=\dv{\tau}{t}|_{t_0}$, $\hat\varphi_{f_0}=\hat\varphi_f(\mathsf{x}(\tau_0))$ and $\langle\hat{O}\rangle_\beta=tr(\hat{O}\sigma_\varphi)$ (see \cite{NK:DM:MM} more detailed calculations).
For a detector which is diagonal in its energy basis simple calculations show that its final purity $r'$ will be equal to
\begin{equation}\label{hot_purity}
r'=re^{-2\lambda^2\dot\tau^2_0\langle\hat\varphi^2_{f_0}\rangle_\beta}\leq r.
\end{equation}
Given that $r=\tanh\frac{\beta_d\Omega}{2}$, with $\beta_d$ an effective inverse temperature, it follows that it is not possible to cool the detector since $\beta_d'\leq\beta_d$. This implies that
\begin{nogotheorem}
It is impossible to construct a relativistic quantum thermal engine using only single instantaneous interactions between the working medium and the reservoirs.
\end{nogotheorem}

It is interesting to note that a similar no-go theorem also exists in the case of entanglement harvesting \cite{Chlimitz:nogo}.

\subsection{Multiple instantaneous interactions}
In order to cool the detector at least two instantaneous interactions are needed
\begin{equation}\label{smeared_interaction}
    \hat{H}_{\text{int}}(\tau)=\sum_{i=1,2}\lambda\delta(t-t_i)\hat{\mu}(\tau)\otimes\hat{\varphi}_f(\mathsf{x}(\tau)).
\end{equation}
In this case it can be shown that the purity of the final state of the detector is of the form (Appendix \ref{first_appendix})
\begin{equation}\label{cold_purity}
r'=Ar+B,
\end{equation}
where	
\begin{align}
A =& e^{-2\lambda^2\dot\tau_2^2\langle\varphi_{f_2}^2\rangle_{\beta}}e^{-2\lambda^2\dot\tau_1^2\langle\varphi_{f_1}^2\rangle_{\beta}}\left(\cos^2\frac{\Omega\Delta\tau}{2}e^{-4\lambda^2\dot\tau_1\dot\tau_2\Re W}\right.\nonumber\\
&\qquad\qquad\qquad\left.+\sin^2\frac{\Omega\Delta\tau}{2}e^{4\lambda^2\dot\tau_1\dot\tau_2\Re W}\right),
\end{align}
\begin{equation}
B = e^{-2\lambda^2\dot\tau_2^2\langle{\varphi_{f_1}^2}\rangle_{\beta}}\sin(\Omega\Delta \tau)\sin({4\lambda^2\dot\tau_1\dot\tau_2\Im W}),
\end{equation}
with $\Delta\tau=\tau_2-\tau_1$, and
\begin{equation}
W=\expval{\hat{\varphi}_f(\mathsf{x}(\tau_1))\hat{\varphi}_f(\mathsf{x}(\tau_2))}_\beta
\end{equation}
is the Wightman function for a smeared thermal field.
\section{The relativistic Otto engine}
\begin{figure*}
\centering
\includegraphics[width=0.8\textwidth]{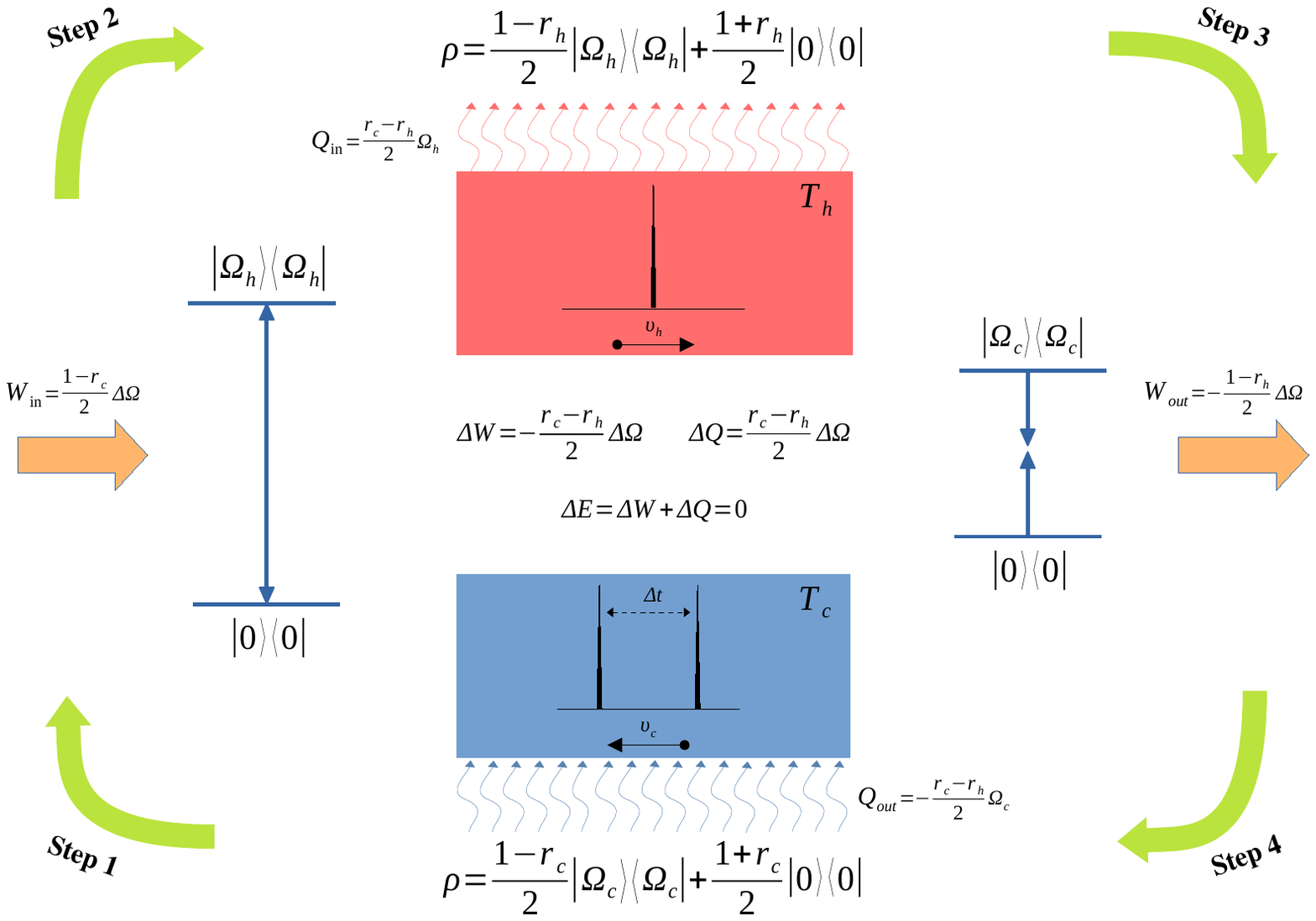}
\caption{The relativistic quantum Otto engine with instantaneous interactions.}
\label{rel_Otto}
\end{figure*}

The relativistic Otto engine consists of the following 4-step process (Fig. \ref{rel_Otto})
\begin{itemize}
    \item[]\textbf{Step 1:} Adiabatic expansion in which the energy gap of the detector is increased from a cold ($\Omega_c$) to a hot ($\Omega_h$) value at the cost of $W_{in}=\frac{1-r_c}{2}\Delta\Omega$ units of work.
    \item[]\textbf{Step 2:} Isochoric contact with a thermal field at a hot temperature $T_h$, through use of a single instantaneous interaction, which changes the purity of the detector from $r_c$ to $r_h$ and draws $Q_{in}=\frac{r_c-r_h}{2}\Omega_h$ units of heat in the process.
    \item[]\textbf{Step 3:} Adiabatic compression which changes the energy gap of the detector back to $\Omega_c$ and releases $W_{out}=-\frac{1-r_h}{2}\Delta\Omega$ units of work.
    \item[]\textbf{Step 4:} Isochoric contact with a thermal field at a cold temperature $T_c$, with the help of two instantaneous interactions, which restores the purity of the detector back to its initial value $r_c$ by dumping $Q_{out}=-\frac{r_c-r_h}{2}\Omega_c$ units of heat.
\end{itemize}
\begin{figure*}
\centering
\includegraphics[width=0.8\textwidth]{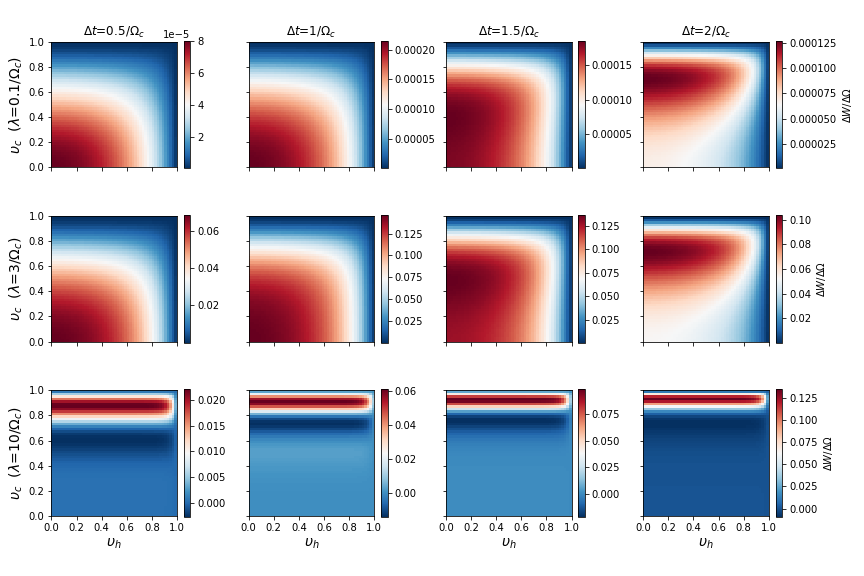}
\caption{Amount of work $\Delta W/\Delta\Omega$ extracted from a relativistic quantum Otto engine as a function of the speed of the detector through the hot ($\upsilon_h$) and cold ($\upsilon_c$) reservoirs for weak ($\lambda_{c,h}=0.1/\Omega_c$), medium ($\lambda_{c,h}=3/\Omega_c$) and strong ($\lambda_{c,h}=10/\Omega_c$) couplings, for a hot $T_h=\Omega_c$ and cold $T_c=0.01\Omega_c$ temperature of the baths and a detector with effective size $R=1/\Omega_c$.}
\label{Otto_field}
\end{figure*}
Adding all the contributions from each step of the cycle, one finds that the total amount of work $\Delta W$ and heat $\Delta Q$ in this case is given by
\begin{equation}\label{work}
\Delta W=-\frac{r_c-r_h}{2}\Delta\Omega\quad\mbox{and}\quad\Delta Q=\frac{r_c-r_h}{2}\Delta\Omega
\end{equation}
with the total energy $\Delta E=\Delta W+\Delta Q=0$, as should be expected from a closed system. If the initial purity of the detector is greater than the one after the second step ($r_c>r_h$) the machine produces useful work at the expense of heat.

Combining Eq. (\ref{hot_purity}) with Eq. (\ref{cold_purity}) it is now possible to find a closed expression for the range of permissable values of the initial purity which provides the necessary and sufficient condition needed in order to be able to close the cycle and repeat the process
\begin{equation}\label{closed_cycle}
r_c=\frac{B}{1-Ae^{-2\lambda^2\dot\tau^2_0\langle\hat\varphi^2_{f_0}\rangle_\beta}},
\end{equation}
with the amount of extracted work now equal to
\begin{equation}
\Delta W=\frac{r_c}{2}(1-e^{-2\lambda^2\dot\tau^2_0\langle\hat\varphi^2_{f_0}\rangle_\beta})\Delta\Omega.
\end{equation}
\subsection{Effects of inertial motion on the amount of extractable work}
We will now study the effects that an inertial detector has on the amount of work that can be extracted from the machine when it is moving through the hot and cold reservoir. The wordline trajectory of the detector in this case is equal to
\begin{equation}
\mathsf{x}(\tau)=\gamma\tau(1,\boldsymbol{\upsilon}_{c,h}),
\end{equation}
where $\gamma=1/\sqrt{1-\upsilon_{c,h}^2}$ is the Lorentz factor and $\upsilon_{c,h}$ denotes the speed of the detector with respect to the cold and hot baths respectively. For a detector with a Gaussian smearing function
\begin{equation}
f(\boldsymbol\xi)=\frac{e^{-\frac{4\abs{{\boldsymbol\xi}}^2}{\pi R^2}}}{(\pi R/2)^3}
\end{equation}
the smeared field takes the form \cite{PhysRevD.105.025006}
\begin{equation}
    \hat\varphi_f=\int\frac{e^{-\frac{\pi k^2\gamma^2R^2(1-\boldsymbol{\upsilon}\cdot\hat\kk)^2}{16}}}{\sqrt{(2\pi)^32\abs{\kk}}}\left(\hat{a}_{\kk}e^{-ik\gamma\tau(1-\boldsymbol\upsilon\cdot\hat\kk)}+\text{H.c.}\right)d^3\kk.
\end{equation}
The expectation values for the field, in this case, can all be written as integrals of error functions, that is
\begin{widetext}
\begin{equation}
\langle\hat\varphi_f^2\rangle_{\beta}=\frac{1}{\sqrt{32\pi^4\gamma^2\upsilon^2R^2}}\int_0^\infty\coth\frac{\beta k}{2}\left[\erf{\sqrt{\frac{\pi}{8}}\gamma kR(1+\upsilon)}-\erf{\sqrt{\frac{\pi}{8}}\gamma kR(1-\upsilon)}\right]dk,
\end{equation}
\begin{equation}
\Re \langle\hat\varphi_{f_1}\hat\varphi_{f_2}\rangle_{\beta}=\frac{e^{-\frac{2\Delta\tau^2}{\pi R^2}}}{\sqrt{32\pi^4\gamma^2\upsilon^2R^2}}\int_0^\infty\coth\frac{\beta k}{2}\left[\Re \erf{\left(\sqrt{\frac{\pi}{8}}\gamma kR(1+\upsilon)+i\sqrt{\frac{2}{\pi}}\frac{\Delta\tau}{R}\right)}-\Re \erf{\left(\sqrt{\frac{\pi}{8}}\gamma kR(1-\upsilon)+i\sqrt{\frac{2}{\pi}}\frac{\Delta\tau}{R}\right)}\right]dk,
\end{equation}
\begin{equation}\label{imaginary}
\Im\langle\hat\varphi_{f_1}\hat\varphi_{f_2}\rangle_{\beta}=\frac{e^{-\frac{2\Delta\tau^2}{\pi R^2}}}{\sqrt{32\pi^4\gamma^2\upsilon^2R^2}}\int_0^\infty\left[\Im\erf{\left(\sqrt{\frac{\pi}{8}}\gamma kR(1-\upsilon)+i\sqrt{\frac{2}{\pi}}\frac{\Delta\tau}{R}\right)}-\Im \erf{\left(\sqrt{\frac{\pi}{8}}\gamma kR(1+\upsilon)+i\sqrt{\frac{2}{\pi}}\frac{\Delta\tau}{R}\right)}\right]dk.
\end{equation}
\end{widetext}

\begin{figure}
\centering
\includegraphics[width=\columnwidth]{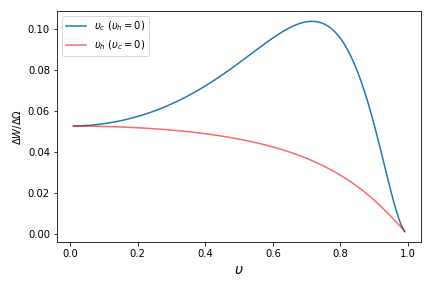}
\caption{Dependence of extracted work on the speed of the detector when it is traveling through the hot and cold heat baths with temperatures $T_h=\Omega_c$ and $T_c=0.01\Omega_c$, for a medium coupling $\lambda_c=\lambda_h=3/\Omega$ and a detector with an effective size $R=1/\Omega_c$. The temporal separation between the two instantaneous interactions used to cool the detector is equal to $\Delta t=2/\Omega$.}
\label{veloc_compare}
\end{figure}

\begin{figure}
\centering
\includegraphics[width=\columnwidth]{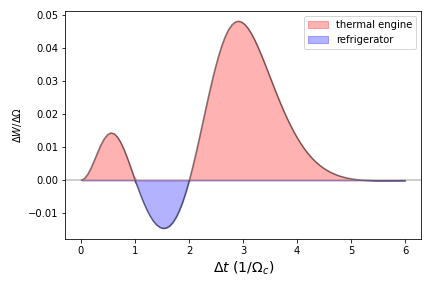}
\caption{Amount of work extracted or expended by a relativistic quantum Otto engine as a function of the temporal separation between the two instantaneous interactions needed to cool the detector for a strong coupling $\lambda_c=\lambda_h=10/\Omega_c$ with the hot and cold reservoir at temperature $T_h=\Omega_c$ and $T_c=0.01\Omega_c$ respectively and for a detector with an effective size $R=1/\Omega_c$. When the work is negative the engine functions as a refrigerator.}
\label{cooling}
\end{figure}

In Fig. \ref{Otto_field}, we present numerical calculations of the amount of work extracted from the Otto engine for a detector with an effective radius $R=1/\Omega_c$, as a function of the speed of the detector $\upsilon_h, \upsilon_c$ through the hot and cold  reservoirs with temperatures $T_c=0.01\Omega_c$ and $T_h=\Omega_c$ respectively, and for various values of the field-detector coupling. We observe that, for the range of parameters considered here, the maximum amount of work that we can recover is more than $25\%$ of the theoretical upper bound ($\Delta W/\Delta\Omega\leq 0.5$). What is striking is the dependence of the work on the speed of the detector through the cold bath. As has been previously reported in \cite{papadatos} the speed of the detector when it is interacting with the hot bath has a degrading effect on the amount of extractable work. On the contrary there exists a non-monotonic relation between work and speed in the cold bath (FIG. \ref{veloc_compare}). Although the physical reason behind this benefit is not immediately clear from the expressions, a clue is given by the persistence of the phenomenon even in the case of strong couplings. In this case $r_c\simeq B$, and the non-monotonic behaviour is a result of the imaginary part of the Wightman function which appears inside a sinusoidal function. As is evident from the figures the maximum contribution to the work occurs at relativistic speeds. In this limit the integral in Eq. (\ref{imaginary}) is dominated by the blue shifted wavelengths due to the relativistic Doppler effect. The moving detector is able to probe modes of the field with a higher energy, which assists in the extraction process. When the couplings with the baths are strong, $r_c$ can take negative values. In this case the machine functions as a refrigerator, drawing heat from the cold bath and dumping it in the hot reservoir at the expense of work done on the detector but with a very low performance no more than $3\%$ of the theoretical bound (see FIG. \ref{cooling}).
\section{Conclusions}
In this report, we demonstrated how to construct a relativistic quantum Otto heat engine using only instantaneous interactions for the isochoric thermal contact between the detector and the heat baths. Employing this approach, which allows for an exact solution of the state of the detector in each stroke of the cycle, we studied the effects of the detector's motion on the amount of work that can be extracted and observed that even though an inertial motion through the hot reservoir tends to degrade work the same motion through the cold reservoir can sometimes enhance the performance of the machine. Numerical calculations indicate that this phenomenon, which occurs for any coupling strength, emerges only for a detector with a size the same order as its inverse energy gap $1/\Omega_c$. Quick calculations show that if the temporal separation between the two instantaneous interactions, needed in the final step of the process in order to cool the detector, is equal to $\Delta t=2/\Omega_c$, then setting $\Delta t=\upsilon_c/\Delta x$ we find that the locations of the two instantaneous interactions in the cold bath are separated by $\Delta x=2\upsilon_c/\Omega_c$ which is the same order as the wavelength of the excited energy level of the detector. This is an ideal scenario for a possible tabletop experiment where a detector could be realized as a two level Fock qubit interacting with an electromagnetic field with a pulse width much shorter than a characteristic length \cite{PhysRevD.105.056023}.

{\bf Note:} During preparation of this manuscript we became aware of a recent similar work \cite{gallockyoshimura2023relativistic}, where a double instantaneous interaction is also employed to construct the Otto engine, but where the isochoric strokes involve the same quantum field instead of two separate fields in a hot and cold temperature. The author also considers the conditions for closing the cycle for a general state of the field in a globally hyperbolic curved spacetime, providing an example of a closed cycle in the case of a static detector that extracts work from a field in the Minkowski vacuum.

\acknowledgments{}
N.K.K. wishes to thank Michał Horodecki, Paweł Horodecki and Paweł Mazurek for fruitful discussions during a visit to The International Centre for Theory of Quantum Technologies in Gdansk where part of this work was presented.
\appendix
\section{UDW detector model with mutliple instantaneous interactions}\label{first_appendix}
For a UDW interaction Hamiltonian written as a sum of delta functions 
\begin{equation}\label{smeared_interaction}
    \hat{H}_{\text{int}}(\tau)=\sum_{i}\lambda_i\delta(t-t_i)\hat{\mu}(\tau)\otimes\hat{\varphi}_f(\mathsf{x}(\tau))
\end{equation}
it can be proven that the unitary operator which evolves the combined system of detector and field
\begin{equation}\label{evolution}
    \hat{U}=\mathcal{T}\text{exp}\left(-i\int\limits_{-\infty}^{+\infty}\hat{H}_{\text{int}}(\tau)d\tau\right),
\end{equation}
where $\mathcal{T}$ is the time-ordering operator, can be written as a product of delta interactions\cite{delta2,pologomez2023nonperturbative} $\hat{U}=\mathcal{T}\prod_i\hat{U}_i$ with
\begin{equation}
\hat{U}_i=P_i^-\otimes V_i+P_i^+\otimes V_i^\dagger
\end{equation}
where $P_i^{\pm}=\frac{I\pm \hat\mu(\tau_i)}{2}$ are projection operators in the detector's Hilbert space and $V_i=e^{i\lambda_i\dot\tau_i\varphi_f(\mathsf{x}(\tau_i))}$  is a unitary rotation acting on the field, with $\tau_i$ the detectors proper time at time $t_i$ and $\dot\tau_i=\dv{\tau}{t}|_{t_i}$.

Restricting our attention to the case of two delta interactions (where without loss of generality we assume that $\lambda_1=\lambda_2=\lambda$) the final state of the detector after the field degrees of freedom have been traced out is given by
\begin{align}
&\Phi(\rho)=tr(\hat{U}\rho\otimes\sigma_\varphi\hat{U}^\dagger)\nonumber\\
&=\frac 12\Pi_2\circ\Pi_1(\rho)+\Pi_2(P_1^-\rho P_1^+)\langle V_1^2\rangle_{\sigma_\varphi}\nonumber\\
&+P_2^-\left(P_1^-\rho P_1^-\langle V_1^\dagger V_2^2 V_1\rangle_{\sigma_\varphi}+P_1^-\rho P_1^+\langle V_1 V_2^2 V_1\rangle_{\sigma_\varphi}\right)P_2^+\nonumber\\
&+P_2^-\left(P_1^+\rho P_1^-\langle V_1^\dagger V_2^2 V_1^\dagger\rangle_{\sigma_\varphi}+P_1^+\rho P_1^+\langle V_1 V_2^2 V_1^\dagger\rangle_{\sigma_\varphi}\right)P_2^+\nonumber\\
&+H.c.
\end{align}
where $\Pi_i(\rho)=P_i^+\rho P_i^++P_i^-\rho P_i^-$. For a thermal field $\sigma_\varphi\propto e^{-\beta\hat{H}_\varphi}$ \cite{NK:DM:MM},
\begin{align}
    &\langle V_1^2\rangle_{\sigma_\varphi}=e^{-2\lambda^2\dot\tau_1^2\expval{\hat{\varphi}_{f_1}^2}_\beta}\nonumber\\
    &\langle V_1^\dagger V_2^2 V_1\rangle_{\sigma_\varphi}=e^{2\lambda^2\dot\tau_1\dot\tau_2[\hat{\varphi}_{f_1},\hat{\varphi}_{f_2}]}e^{-2\lambda^2\dot\tau_2^2\expval{\hat{\varphi}_{f_2}^2}_\beta}\nonumber\\
    &\langle V_1 V_2^2 V_1^\dagger\rangle_{\sigma_\varphi}=e^{-2\lambda^2[\hat{\varphi}_{f_1},\hat{\varphi}_{f_2}]}e^{-2\lambda^2\dot\tau_2^2\expval{\hat{\varphi}_{f_2}^2}_\beta}\\
    &\langle V_1^\dagger V_2^2 V_1^\dagger\rangle_{\sigma_\varphi}=e^{-2\lambda^2\expval{(\dot\tau_2\hat{\varphi}_{f_2}-\dot\tau_1\hat{\varphi}_{f_1})^2}_\beta}\nonumber\\
    &\langle V_1 V_2^2 V_1\rangle_{\sigma_\varphi}=e^{-2\lambda^2\expval{(\dot\tau_2\hat{\varphi}_{f_2}+\dot\tau_1\hat{\varphi}_{f_1})^2}_\beta}\nonumber
\end{align}
 For a detector with initial state $\rho=\frac{1}{2}(I-rZ)$ where $Z$ denotes the z-Pauli matrix
 \begin{equation}
 \Phi(\rho)=\frac{1}{2}(I-r'Z)
 \end{equation}
 where $r'$ is given in Eq. (\ref{cold_purity}).
\bibliography{references}

\begin{thebibliography}{61}%
\makeatletter
\providecommand \@ifxundefined [1]{%
 \@ifx{#1\undefined}
}%
\providecommand \@ifnum [1]{%
 \ifnum #1\expandafter \@firstoftwo
 \else \expandafter \@secondoftwo
 \fi
}%
\providecommand \@ifx [1]{%
 \ifx #1\expandafter \@firstoftwo
 \else \expandafter \@secondoftwo
 \fi
}%
\providecommand \natexlab [1]{#1}%
\providecommand \enquote  [1]{``#1''}%
\providecommand \bibnamefont  [1]{#1}%
\providecommand \bibfnamefont [1]{#1}%
\providecommand \citenamefont [1]{#1}%
\providecommand \href@noop [0]{\@secondoftwo}%
\providecommand \href [0]{\begingroup \@sanitize@url \@href}%
\providecommand \@href[1]{\@@startlink{#1}\@@href}%
\providecommand \@@href[1]{\endgroup#1\@@endlink}%
\providecommand \@sanitize@url [0]{\catcode `\\12\catcode `\$12\catcode `\&12\catcode `\#12\catcode `\^12\catcode `\_12\catcode `\%12\relax}%
\providecommand \@@startlink[1]{}%
\providecommand \@@endlink[0]{}%
\providecommand \url  [0]{\begingroup\@sanitize@url \@url }%
\providecommand \@url [1]{\endgroup\@href {#1}{\urlprefix }}%
\providecommand \urlprefix  [0]{URL }%
\providecommand \Eprint [0]{\href }%
\providecommand \doibase [0]{http://dx.doi.org/}%
\providecommand \selectlanguage [0]{\@gobble}%
\providecommand \bibinfo  [0]{\@secondoftwo}%
\providecommand \bibfield  [0]{\@secondoftwo}%
\providecommand \translation [1]{[#1]}%
\providecommand \BibitemOpen [0]{}%
\providecommand \bibitemStop [0]{}%
\providecommand \bibitemNoStop [0]{.\EOS\space}%
\providecommand \EOS [0]{\spacefactor3000\relax}%
\providecommand \BibitemShut  [1]{\csname bibitem#1\endcsname}%
\let\auto@bib@innerbib\@empty
%</preamble>
\bibitem [{\citenamefont {Goold}\ \emph {et~al.}(2016)\citenamefont {Goold}, \citenamefont {Huber}, \citenamefont {Riera}, \citenamefont {Del~Rio},\ and\ \citenamefont {Skrzypczyk}}]{goold2016role}%
  \BibitemOpen
  \bibfield  {author} {\bibinfo {author} {\bibfnamefont {John}\ \bibnamefont {Goold}}, \bibinfo {author} {\bibfnamefont {Marcus}\ \bibnamefont {Huber}}, \bibinfo {author} {\bibfnamefont {Arnau}\ \bibnamefont {Riera}}, \bibinfo {author} {\bibfnamefont {L{\'\i}dia}\ \bibnamefont {Del~Rio}}, \ and\ \bibinfo {author} {\bibfnamefont {Paul}\ \bibnamefont {Skrzypczyk}},\ }\bibfield  {title} {\enquote {\bibinfo {title} {The role of quantum information in thermodynamics—a topical review},}\ }\href@noop {} {\bibfield  {journal} {\bibinfo  {journal} {Journal of Physics A: Mathematical and Theoretical}\ }\textbf {\bibinfo {volume} {49}},\ \bibinfo {pages} {143001} (\bibinfo {year} {2016})}\BibitemShut {NoStop}%
\bibitem [{\citenamefont {Vinjanampathy}\ and\ \citenamefont {Anders}(2016)}]{vinjanampathy2016quantum}%
  \BibitemOpen
  \bibfield  {author} {\bibinfo {author} {\bibfnamefont {Sai}\ \bibnamefont {Vinjanampathy}}\ and\ \bibinfo {author} {\bibfnamefont {Janet}\ \bibnamefont {Anders}},\ }\bibfield  {title} {\enquote {\bibinfo {title} {Quantum thermodynamics},}\ }\href@noop {} {\bibfield  {journal} {\bibinfo  {journal} {Contemporary Physics}\ }\textbf {\bibinfo {volume} {57}},\ \bibinfo {pages} {545--579} (\bibinfo {year} {2016})}\BibitemShut {NoStop}%
\bibitem [{\citenamefont {{Lostaglio}}(2019)}]{Lostaglio}%
  \BibitemOpen
  \bibfield  {author} {\bibinfo {author} {\bibfnamefont {Matteo}\ \bibnamefont {{Lostaglio}}},\ }\bibfield  {title} {\enquote {\bibinfo {title} {{An introductory review of the resource theory approach to thermodynamics}},}\ }\href {\doibase 10.1088/1361-6633/ab46e5} {\bibfield  {journal} {\bibinfo  {journal} {Reports on Progress in Physics}\ }\textbf {\bibinfo {volume} {82}},\ \bibinfo {eid} {114001} (\bibinfo {year} {2019})}\BibitemShut {NoStop}%
\bibitem [{\citenamefont {Binder}\ \emph {et~al.}(2018)\citenamefont {Binder}, \citenamefont {Correa}, \citenamefont {Gogolin}, \citenamefont {Anders},\ and\ \citenamefont {Adesso}}]{binder2019thermodynamics}%
  \BibitemOpen
  \bibfield  {author} {\bibinfo {author} {\bibfnamefont {F.}~\bibnamefont {Binder}}, \bibinfo {author} {\bibfnamefont {L.A.}\ \bibnamefont {Correa}}, \bibinfo {author} {\bibfnamefont {C.}~\bibnamefont {Gogolin}}, \bibinfo {author} {\bibfnamefont {J.}~\bibnamefont {Anders}}, \ and\ \bibinfo {author} {\bibfnamefont {G.}~\bibnamefont {Adesso}},\ }\href@noop {} {\emph {\bibinfo {title} {Thermodynamics in the Quantum Regime: Fundamental Aspects and New Directions}}},\ Fundamental Theories of Physics\ (\bibinfo  {publisher} {Springer International Publishing},\ \bibinfo {year} {2018})\BibitemShut {NoStop}%
\bibitem [{\citenamefont {Deffner}\ and\ \citenamefont {Campbell}(2019)}]{Deffner:Campell}%
  \BibitemOpen
  \bibfield  {author} {\bibinfo {author} {\bibfnamefont {Sebastian}\ \bibnamefont {Deffner}}\ and\ \bibinfo {author} {\bibfnamefont {Steve}\ \bibnamefont {Campbell}},\ }\href {\doibase 10.1088/2053-2571/ab21c6} {\emph {\bibinfo {title} {Quantum Thermodynamics}}},\ 2053-2571\ (\bibinfo  {publisher} {Morgan \& Claypool Publishers},\ \bibinfo {year} {2019})\BibitemShut {NoStop}%
\bibitem [{\citenamefont {{Kosloff}}\ and\ \citenamefont {{Levy}}(2014)}]{Kosloff}%
  \BibitemOpen
  \bibfield  {author} {\bibinfo {author} {\bibfnamefont {Ronnie}\ \bibnamefont {{Kosloff}}}\ and\ \bibinfo {author} {\bibfnamefont {Amikam}\ \bibnamefont {{Levy}}},\ }\bibfield  {title} {\enquote {\bibinfo {title} {{Quantum Heat Engines and Refrigerators: Continuous Devices}},}\ }\href {\doibase 10.1146/annurev-physchem-040513-103724} {\bibfield  {journal} {\bibinfo  {journal} {Annual Review of Physical Chemistry}\ }\textbf {\bibinfo {volume} {65}},\ \bibinfo {pages} {365--393} (\bibinfo {year} {2014})}\BibitemShut {NoStop}%
\bibitem [{\citenamefont {{Myers}}\ \emph {et~al.}(2022)\citenamefont {{Myers}}, \citenamefont {{Abah}},\ and\ \citenamefont {{Deffner}}}]{Deffner:TMs}%
  \BibitemOpen
  \bibfield  {author} {\bibinfo {author} {\bibfnamefont {Nathan~M.}\ \bibnamefont {{Myers}}}, \bibinfo {author} {\bibfnamefont {Obinna}\ \bibnamefont {{Abah}}}, \ and\ \bibinfo {author} {\bibfnamefont {Sebastian}\ \bibnamefont {{Deffner}}},\ }\bibfield  {title} {\enquote {\bibinfo {title} {{Quantum thermodynamic devices: From theoretical proposals to experimental reality}},}\ }\href {\doibase 10.1116/5.0083192} {\bibfield  {journal} {\bibinfo  {journal} {AVS Quantum Science}\ }\textbf {\bibinfo {volume} {4}},\ \bibinfo {eid} {027101} (\bibinfo {year} {2022})}\BibitemShut {NoStop}%
\bibitem [{\citenamefont {Bhattacharjee}\ and\ \citenamefont {Dutta}(2021)}]{TMs:and:batt}%
  \BibitemOpen
  \bibfield  {author} {\bibinfo {author} {\bibfnamefont {Sourav}\ \bibnamefont {Bhattacharjee}}\ and\ \bibinfo {author} {\bibfnamefont {Amit}\ \bibnamefont {Dutta}},\ }\bibfield  {title} {\enquote {\bibinfo {title} {Quantum thermal machines and batteries},}\ }\href {https://link.springer.com/article/10.1140/epjb/s10051-021-00235-3} {\bibfield  {journal} {\bibinfo  {journal} {The European Physical Journal B}\ }\textbf {\bibinfo {volume} {94}},\ \bibinfo {pages} {239} (\bibinfo {year} {2021})}\BibitemShut {NoStop}%
\bibitem [{\citenamefont {Cangemi}\ \emph {et~al.}(2023)\citenamefont {Cangemi}, \citenamefont {Bhadra},\ and\ \citenamefont {Levy}}]{cangemi2023quantum}%
  \BibitemOpen
  \bibfield  {author} {\bibinfo {author} {\bibfnamefont {Loris~Maria}\ \bibnamefont {Cangemi}}, \bibinfo {author} {\bibfnamefont {Chitrak}\ \bibnamefont {Bhadra}}, \ and\ \bibinfo {author} {\bibfnamefont {Amikam}\ \bibnamefont {Levy}},\ }\href@noop {} {\enquote {\bibinfo {title} {Quantum engines and refrigerators},}\ } (\bibinfo {year} {2023}),\ \Eprint {http://arxiv.org/abs/2302.00726} {arXiv:2302.00726 [quant-ph]} \BibitemShut {NoStop}%
\bibitem [{\citenamefont {Scovil}\ and\ \citenamefont {Schulz-DuBois}(1959)}]{Scovil}%
  \BibitemOpen
  \bibfield  {author} {\bibinfo {author} {\bibfnamefont {H.~E.~D.}\ \bibnamefont {Scovil}}\ and\ \bibinfo {author} {\bibfnamefont {E.~O.}\ \bibnamefont {Schulz-DuBois}},\ }\bibfield  {title} {\enquote {\bibinfo {title} {Three-level masers as heat engines},}\ }\href {\doibase 10.1103/PhysRevLett.2.262} {\bibfield  {journal} {\bibinfo  {journal} {Phys. Rev. Lett.}\ }\textbf {\bibinfo {volume} {2}},\ \bibinfo {pages} {262--263} (\bibinfo {year} {1959})}\BibitemShut {NoStop}%
\bibitem [{\citenamefont {Scully}\ \emph {et~al.}(2003)\citenamefont {Scully}, \citenamefont {Zubairy}, \citenamefont {Agarwal},\ and\ \citenamefont {Walther}}]{Scully:Coh}%
  \BibitemOpen
  \bibfield  {author} {\bibinfo {author} {\bibfnamefont {Marlan~O.}\ \bibnamefont {Scully}}, \bibinfo {author} {\bibfnamefont {M.~Suhail}\ \bibnamefont {Zubairy}}, \bibinfo {author} {\bibfnamefont {Girish~S.}\ \bibnamefont {Agarwal}}, \ and\ \bibinfo {author} {\bibfnamefont {Herbert}\ \bibnamefont {Walther}},\ }\bibfield  {title} {\enquote {\bibinfo {title} {Extracting work from a single heat bath via vanishing quantum coherence},}\ }\href {\doibase 10.1126/science.1078955} {\bibfield  {journal} {\bibinfo  {journal} {Science}\ }\textbf {\bibinfo {volume} {299}},\ \bibinfo {pages} {862--864} (\bibinfo {year} {2003})}\BibitemShut {NoStop}%
\bibitem [{\citenamefont {Uzdin}\ \emph {et~al.}(2015)\citenamefont {Uzdin}, \citenamefont {Levy},\ and\ \citenamefont {Kosloff}}]{QTM:coh}%
  \BibitemOpen
  \bibfield  {author} {\bibinfo {author} {\bibfnamefont {Raam}\ \bibnamefont {Uzdin}}, \bibinfo {author} {\bibfnamefont {Amikam}\ \bibnamefont {Levy}}, \ and\ \bibinfo {author} {\bibfnamefont {Ronnie}\ \bibnamefont {Kosloff}},\ }\bibfield  {title} {\enquote {\bibinfo {title} {Equivalence of quantum heat machines, and quantum-thermodynamic signatures},}\ }\href {\doibase 10.1103/PhysRevX.5.031044} {\bibfield  {journal} {\bibinfo  {journal} {Phys. Rev. X}\ }\textbf {\bibinfo {volume} {5}},\ \bibinfo {pages} {031044} (\bibinfo {year} {2015})}\BibitemShut {NoStop}%
\bibitem [{\citenamefont {Ro\ss{}nagel}\ \emph {et~al.}(2014)\citenamefont {Ro\ss{}nagel}, \citenamefont {Abah}, \citenamefont {Schmidt-Kaler}, \citenamefont {Singer},\ and\ \citenamefont {Lutz}}]{QTM:squeez}%
  \BibitemOpen
  \bibfield  {author} {\bibinfo {author} {\bibfnamefont {J.}~\bibnamefont {Ro\ss{}nagel}}, \bibinfo {author} {\bibfnamefont {O.}~\bibnamefont {Abah}}, \bibinfo {author} {\bibfnamefont {F.}~\bibnamefont {Schmidt-Kaler}}, \bibinfo {author} {\bibfnamefont {K.}~\bibnamefont {Singer}}, \ and\ \bibinfo {author} {\bibfnamefont {E.}~\bibnamefont {Lutz}},\ }\bibfield  {title} {\enquote {\bibinfo {title} {Nanoscale heat engine beyond the carnot limit},}\ }\href {\doibase 10.1103/PhysRevLett.112.030602} {\bibfield  {journal} {\bibinfo  {journal} {Phys. Rev. Lett.}\ }\textbf {\bibinfo {volume} {112}},\ \bibinfo {pages} {030602} (\bibinfo {year} {2014})}\BibitemShut {NoStop}%
\bibitem [{\citenamefont {Agarwalla}\ \emph {et~al.}(2017)\citenamefont {Agarwalla}, \citenamefont {Jiang},\ and\ \citenamefont {Segal}}]{QTM:squeez:2}%
  \BibitemOpen
  \bibfield  {author} {\bibinfo {author} {\bibfnamefont {Bijay~Kumar}\ \bibnamefont {Agarwalla}}, \bibinfo {author} {\bibfnamefont {Jian-Hua}\ \bibnamefont {Jiang}}, \ and\ \bibinfo {author} {\bibfnamefont {Dvira}\ \bibnamefont {Segal}},\ }\bibfield  {title} {\enquote {\bibinfo {title} {Quantum efficiency bound for continuous heat engines coupled to noncanonical reservoirs},}\ }\href {\doibase 10.1103/PhysRevB.96.104304} {\bibfield  {journal} {\bibinfo  {journal} {Phys. Rev. B}\ }\textbf {\bibinfo {volume} {96}},\ \bibinfo {pages} {104304} (\bibinfo {year} {2017})}\BibitemShut {NoStop}%
\bibitem [{\citenamefont {{Niedenzu}}\ \emph {et~al.}(2018)\citenamefont {{Niedenzu}}, \citenamefont {{Mukherjee}}, \citenamefont {{Ghosh}}, \citenamefont {{Kofman}},\ and\ \citenamefont {{Kurizki}}}]{QTM:squeez:3}%
  \BibitemOpen
  \bibfield  {author} {\bibinfo {author} {\bibfnamefont {Wolfgang}\ \bibnamefont {{Niedenzu}}}, \bibinfo {author} {\bibfnamefont {Victor}\ \bibnamefont {{Mukherjee}}}, \bibinfo {author} {\bibfnamefont {Arnab}\ \bibnamefont {{Ghosh}}}, \bibinfo {author} {\bibfnamefont {Abraham~G.}\ \bibnamefont {{Kofman}}}, \ and\ \bibinfo {author} {\bibfnamefont {Gershon}\ \bibnamefont {{Kurizki}}},\ }\bibfield  {title} {\enquote {\bibinfo {title} {{Quantum engine efficiency bound beyond the second law of thermodynamics}},}\ }\href {\doibase 10.1038/s41467-017-01991-6} {\bibfield  {journal} {\bibinfo  {journal} {Nature Communications}\ }\textbf {\bibinfo {volume} {9}},\ \bibinfo {eid} {165} (\bibinfo {year} {2018})}\BibitemShut {NoStop}%
\bibitem [{\citenamefont {Gelbwaser-Klimovsky}\ \emph {et~al.}(2013)\citenamefont {Gelbwaser-Klimovsky}, \citenamefont {Erez}, \citenamefont {Alicki},\ and\ \citenamefont {Kurizki}}]{Non:Mark:QTM}%
  \BibitemOpen
  \bibfield  {author} {\bibinfo {author} {\bibfnamefont {D.}~\bibnamefont {Gelbwaser-Klimovsky}}, \bibinfo {author} {\bibfnamefont {N.}~\bibnamefont {Erez}}, \bibinfo {author} {\bibfnamefont {R.}~\bibnamefont {Alicki}}, \ and\ \bibinfo {author} {\bibfnamefont {G.}~\bibnamefont {Kurizki}},\ }\bibfield  {title} {\enquote {\bibinfo {title} {Work extraction via quantum nondemolition measurements of qubits in cavities: Non-markovian effects},}\ }\href {\doibase 10.1103/PhysRevA.88.022112} {\bibfield  {journal} {\bibinfo  {journal} {Phys. Rev. A}\ }\textbf {\bibinfo {volume} {88}},\ \bibinfo {pages} {022112} (\bibinfo {year} {2013})}\BibitemShut {NoStop}%
\bibitem [{\citenamefont {{Pezzutto}}\ \emph {et~al.}(2019)\citenamefont {{Pezzutto}}, \citenamefont {{Paternostro}},\ and\ \citenamefont {{Omar}}}]{Non:Mark:QTM:2}%
  \BibitemOpen
  \bibfield  {author} {\bibinfo {author} {\bibfnamefont {Marco}\ \bibnamefont {{Pezzutto}}}, \bibinfo {author} {\bibfnamefont {Mauro}\ \bibnamefont {{Paternostro}}}, \ and\ \bibinfo {author} {\bibfnamefont {Yasser}\ \bibnamefont {{Omar}}},\ }\bibfield  {title} {\enquote {\bibinfo {title} {{An out-of-equilibrium non-Markovian quantum heat engine}},}\ }\href {\doibase 10.1088/2058-9565/aaf5b4} {\bibfield  {journal} {\bibinfo  {journal} {Quantum Science and Technology}\ }\textbf {\bibinfo {volume} {4}},\ \bibinfo {pages} {025002} (\bibinfo {year} {2019})}\BibitemShut {NoStop}%
\bibitem [{\citenamefont {{Strasberg}}\ \emph {et~al.}(2016)\citenamefont {{Strasberg}}, \citenamefont {{Schaller}}, \citenamefont {{Lambert}},\ and\ \citenamefont {{Brandes}}}]{QTM:strong}%
  \BibitemOpen
  \bibfield  {author} {\bibinfo {author} {\bibfnamefont {Philipp}\ \bibnamefont {{Strasberg}}}, \bibinfo {author} {\bibfnamefont {Gernot}\ \bibnamefont {{Schaller}}}, \bibinfo {author} {\bibfnamefont {Neill}\ \bibnamefont {{Lambert}}}, \ and\ \bibinfo {author} {\bibfnamefont {Tobias}\ \bibnamefont {{Brandes}}},\ }\bibfield  {title} {\enquote {\bibinfo {title} {{Nonequilibrium thermodynamics in the strong coupling and non-Markovian regime based on a reaction coordinate mapping}},}\ }\href {\doibase 10.1088/1367-2630/18/7/073007} {\bibfield  {journal} {\bibinfo  {journal} {New Journal of Physics}\ }\textbf {\bibinfo {volume} {18}},\ \bibinfo {eid} {073007} (\bibinfo {year} {2016})}\BibitemShut {NoStop}%
\bibitem [{\citenamefont {Newman}\ \emph {et~al.}(2017)\citenamefont {Newman}, \citenamefont {Mintert},\ and\ \citenamefont {Nazir}}]{QTM:strong:2}%
  \BibitemOpen
  \bibfield  {author} {\bibinfo {author} {\bibfnamefont {David}\ \bibnamefont {Newman}}, \bibinfo {author} {\bibfnamefont {Florian}\ \bibnamefont {Mintert}}, \ and\ \bibinfo {author} {\bibfnamefont {Ahsan}\ \bibnamefont {Nazir}},\ }\bibfield  {title} {\enquote {\bibinfo {title} {Performance of a quantum heat engine at strong reservoir coupling},}\ }\href {\doibase 10.1103/PhysRevE.95.032139} {\bibfield  {journal} {\bibinfo  {journal} {Phys. Rev. E}\ }\textbf {\bibinfo {volume} {95}},\ \bibinfo {pages} {032139} (\bibinfo {year} {2017})}\BibitemShut {NoStop}%
\bibitem [{\citenamefont {Perarnau-Llobet}\ \emph {et~al.}(2018)\citenamefont {Perarnau-Llobet}, \citenamefont {Wilming}, \citenamefont {Riera}, \citenamefont {Gallego},\ and\ \citenamefont {Eisert}}]{QTM:strong:3}%
  \BibitemOpen
  \bibfield  {author} {\bibinfo {author} {\bibfnamefont {M.}~\bibnamefont {Perarnau-Llobet}}, \bibinfo {author} {\bibfnamefont {H.}~\bibnamefont {Wilming}}, \bibinfo {author} {\bibfnamefont {A.}~\bibnamefont {Riera}}, \bibinfo {author} {\bibfnamefont {R.}~\bibnamefont {Gallego}}, \ and\ \bibinfo {author} {\bibfnamefont {J.}~\bibnamefont {Eisert}},\ }\bibfield  {title} {\enquote {\bibinfo {title} {Strong coupling corrections in quantum thermodynamics},}\ }\href {\doibase 10.1103/PhysRevLett.120.120602} {\bibfield  {journal} {\bibinfo  {journal} {Phys. Rev. Lett.}\ }\textbf {\bibinfo {volume} {120}},\ \bibinfo {pages} {120602} (\bibinfo {year} {2018})}\BibitemShut {NoStop}%
\bibitem [{\citenamefont {Abah}\ \emph {et~al.}(2012)\citenamefont {Abah}, \citenamefont {Ro\ss{}nagel}, \citenamefont {Jacob}, \citenamefont {Deffner}, \citenamefont {Schmidt-Kaler}, \citenamefont {Singer},\ and\ \citenamefont {Lutz}}]{exp:eng}%
  \BibitemOpen
  \bibfield  {author} {\bibinfo {author} {\bibfnamefont {O.}~\bibnamefont {Abah}}, \bibinfo {author} {\bibfnamefont {J.}~\bibnamefont {Ro\ss{}nagel}}, \bibinfo {author} {\bibfnamefont {G.}~\bibnamefont {Jacob}}, \bibinfo {author} {\bibfnamefont {S.}~\bibnamefont {Deffner}}, \bibinfo {author} {\bibfnamefont {F.}~\bibnamefont {Schmidt-Kaler}}, \bibinfo {author} {\bibfnamefont {K.}~\bibnamefont {Singer}}, \ and\ \bibinfo {author} {\bibfnamefont {E.}~\bibnamefont {Lutz}},\ }\bibfield  {title} {\enquote {\bibinfo {title} {Single-ion heat engine at maximum power},}\ }\href {\doibase 10.1103/PhysRevLett.109.203006} {\bibfield  {journal} {\bibinfo  {journal} {Phys. Rev. Lett.}\ }\textbf {\bibinfo {volume} {109}},\ \bibinfo {pages} {203006} (\bibinfo {year} {2012})}\BibitemShut {NoStop}%
\bibitem [{\citenamefont {{Ro{\ss}nagel}}\ \emph {et~al.}(2016)\citenamefont {{Ro{\ss}nagel}}, \citenamefont {{Dawkins}}, \citenamefont {{Tolazzi}}, \citenamefont {{Abah}}, \citenamefont {{Lutz}}, \citenamefont {{Schmidt-Kaler}},\ and\ \citenamefont {{Singer}}}]{Exp:eng:single:atom}%
  \BibitemOpen
  \bibfield  {author} {\bibinfo {author} {\bibfnamefont {Johannes}\ \bibnamefont {{Ro{\ss}nagel}}}, \bibinfo {author} {\bibfnamefont {Samuel~T.}\ \bibnamefont {{Dawkins}}}, \bibinfo {author} {\bibfnamefont {Karl~N.}\ \bibnamefont {{Tolazzi}}}, \bibinfo {author} {\bibfnamefont {Obinna}\ \bibnamefont {{Abah}}}, \bibinfo {author} {\bibfnamefont {Eric}\ \bibnamefont {{Lutz}}}, \bibinfo {author} {\bibfnamefont {Ferdinand}\ \bibnamefont {{Schmidt-Kaler}}}, \ and\ \bibinfo {author} {\bibfnamefont {Kilian}\ \bibnamefont {{Singer}}},\ }\bibfield  {title} {\enquote {\bibinfo {title} {{A single-atom heat engine}},}\ }\href {\doibase 10.1126/science.aad6320} {\bibfield  {journal} {\bibinfo  {journal} {Science}\ }\textbf {\bibinfo {volume} {352}},\ \bibinfo {pages} {325--329} (\bibinfo {year} {2016})}\BibitemShut {NoStop}%
\bibitem [{\citenamefont {von Lindenfels}\ \emph {et~al.}(2019)\citenamefont {von Lindenfels}, \citenamefont {Gr\"ab}, \citenamefont {Schmiegelow}, \citenamefont {Kaushal}, \citenamefont {Schulz}, \citenamefont {Mitchison}, \citenamefont {Goold}, \citenamefont {Schmidt-Kaler},\ and\ \citenamefont {Poschinger}}]{Exp:engine:ion:spin}%
  \BibitemOpen
  \bibfield  {author} {\bibinfo {author} {\bibfnamefont {D.}~\bibnamefont {von Lindenfels}}, \bibinfo {author} {\bibfnamefont {O.}~\bibnamefont {Gr\"ab}}, \bibinfo {author} {\bibfnamefont {C.~T.}\ \bibnamefont {Schmiegelow}}, \bibinfo {author} {\bibfnamefont {V.}~\bibnamefont {Kaushal}}, \bibinfo {author} {\bibfnamefont {J.}~\bibnamefont {Schulz}}, \bibinfo {author} {\bibfnamefont {Mark~T.}\ \bibnamefont {Mitchison}}, \bibinfo {author} {\bibfnamefont {John}\ \bibnamefont {Goold}}, \bibinfo {author} {\bibfnamefont {F.}~\bibnamefont {Schmidt-Kaler}}, \ and\ \bibinfo {author} {\bibfnamefont {U.~G.}\ \bibnamefont {Poschinger}},\ }\bibfield  {title} {\enquote {\bibinfo {title} {Spin heat engine coupled to a harmonic-oscillator flywheel},}\ }\href {\doibase 10.1103/PhysRevLett.123.080602} {\bibfield  {journal} {\bibinfo  {journal} {Phys. Rev. Lett.}\ }\textbf {\bibinfo {volume} {123}},\ \bibinfo {pages} {080602} (\bibinfo {year} {2019})}\BibitemShut {NoStop}%
\bibitem [{\citenamefont {Peterson}\ \emph {et~al.}(2019)\citenamefont {Peterson}, \citenamefont {Batalh\~ao}, \citenamefont {Herrera}, \citenamefont {Souza}, \citenamefont {Sarthour}, \citenamefont {Oliveira},\ and\ \citenamefont {Serra}}]{exp:spin:eng}%
  \BibitemOpen
  \bibfield  {author} {\bibinfo {author} {\bibfnamefont {John P.~S.}\ \bibnamefont {Peterson}}, \bibinfo {author} {\bibfnamefont {Tiago~B.}\ \bibnamefont {Batalh\~ao}}, \bibinfo {author} {\bibfnamefont {Marcela}\ \bibnamefont {Herrera}}, \bibinfo {author} {\bibfnamefont {Alexandre~M.}\ \bibnamefont {Souza}}, \bibinfo {author} {\bibfnamefont {Roberto~S.}\ \bibnamefont {Sarthour}}, \bibinfo {author} {\bibfnamefont {Ivan~S.}\ \bibnamefont {Oliveira}}, \ and\ \bibinfo {author} {\bibfnamefont {Roberto~M.}\ \bibnamefont {Serra}},\ }\bibfield  {title} {\enquote {\bibinfo {title} {Experimental characterization of a spin quantum heat engine},}\ }\href {\doibase 10.1103/PhysRevLett.123.240601} {\bibfield  {journal} {\bibinfo  {journal} {Phys. Rev. Lett.}\ }\textbf {\bibinfo {volume} {123}},\ \bibinfo {pages} {240601} (\bibinfo {year} {2019})}\BibitemShut {NoStop}%
\bibitem [{\citenamefont {Gluza}\ \emph {et~al.}(2021)\citenamefont {Gluza}, \citenamefont {Sabino}, \citenamefont {Ng}, \citenamefont {Vitagliano}, \citenamefont {Pezzutto}, \citenamefont {Omar}, \citenamefont {Mazets}, \citenamefont {Huber}, \citenamefont {Schmiedmayer},\ and\ \citenamefont {Eisert}}]{Exp:QFT:machine:Huber}%
  \BibitemOpen
  \bibfield  {author} {\bibinfo {author} {\bibfnamefont {Marek}\ \bibnamefont {Gluza}}, \bibinfo {author} {\bibfnamefont {Jo\~ao}\ \bibnamefont {Sabino}}, \bibinfo {author} {\bibfnamefont {Nelly~H.Y.}\ \bibnamefont {Ng}}, \bibinfo {author} {\bibfnamefont {Giuseppe}\ \bibnamefont {Vitagliano}}, \bibinfo {author} {\bibfnamefont {Marco}\ \bibnamefont {Pezzutto}}, \bibinfo {author} {\bibfnamefont {Yasser}\ \bibnamefont {Omar}}, \bibinfo {author} {\bibfnamefont {Igor}\ \bibnamefont {Mazets}}, \bibinfo {author} {\bibfnamefont {Marcus}\ \bibnamefont {Huber}}, \bibinfo {author} {\bibfnamefont {J\"org}\ \bibnamefont {Schmiedmayer}}, \ and\ \bibinfo {author} {\bibfnamefont {Jens}\ \bibnamefont {Eisert}},\ }\bibfield  {title} {\enquote {\bibinfo {title} {Quantum field thermal machines},}\ }\href {\doibase 10.1103/PRXQuantum.2.030310} {\bibfield  {journal} {\bibinfo  {journal} {PRX Quantum}\ }\textbf {\bibinfo {volume} {2}},\ \bibinfo {pages} {030310} (\bibinfo {year} {2021})}\BibitemShut {NoStop}%
\bibitem [{\citenamefont {Unruh}(1976)}]{Unruh}%
  \BibitemOpen
  \bibfield  {author} {\bibinfo {author} {\bibfnamefont {W.~G.}\ \bibnamefont {Unruh}},\ }\bibfield  {title} {\enquote {\bibinfo {title} {Notes on black-hole evaporation},}\ }\href {\doibase 10.1103/PhysRevD.14.870} {\bibfield  {journal} {\bibinfo  {journal} {Phys. Rev. D}\ }\textbf {\bibinfo {volume} {14}},\ \bibinfo {pages} {870--892} (\bibinfo {year} {1976})}\BibitemShut {NoStop}%
\bibitem [{\citenamefont {Takagi}(1986)}]{Takagi}%
  \BibitemOpen
  \bibfield  {author} {\bibinfo {author} {\bibfnamefont {Shin}\ \bibnamefont {Takagi}},\ }\bibfield  {title} {\enquote {\bibinfo {title} {{Vacuum Noise and Stress Induced by Uniform Acceleration: Hawking-Unruh Effect in Rindler Manifold of Arbitrary Dimension}},}\ }\href {\doibase 10.1143/PTP.88.1} {\bibfield  {journal} {\bibinfo  {journal} {Progress of Theoretical Physics Supplement}\ }\textbf {\bibinfo {volume} {88}},\ \bibinfo {pages} {1--142} (\bibinfo {year} {1986})}\BibitemShut {NoStop}%
\bibitem [{\citenamefont {{Fewster}}\ \emph {et~al.}(2016)\citenamefont {{Fewster}}, \citenamefont {{Ju{\'a}rez-Aubry}},\ and\ \citenamefont {{Louko}}}]{Louko:Fewster}%
  \BibitemOpen
  \bibfield  {author} {\bibinfo {author} {\bibfnamefont {Christopher~J.}\ \bibnamefont {{Fewster}}}, \bibinfo {author} {\bibfnamefont {Benito~A.}\ \bibnamefont {{Ju{\'a}rez-Aubry}}}, \ and\ \bibinfo {author} {\bibfnamefont {Jorma}\ \bibnamefont {{Louko}}},\ }\bibfield  {title} {\enquote {\bibinfo {title} {{Waiting for Unruh}},}\ }\href {\doibase 10.1088/0264-9381/33/16/165003} {\bibfield  {journal} {\bibinfo  {journal} {Classical and Quantum Gravity}\ }\textbf {\bibinfo {volume} {33}},\ \bibinfo {eid} {165003} (\bibinfo {year} {2016})}\BibitemShut {NoStop}%
\bibitem [{\citenamefont {Ju\'arez-Aubry}\ and\ \citenamefont {Moustos}(2019)}]{BJA:DM}%
  \BibitemOpen
  \bibfield  {author} {\bibinfo {author} {\bibfnamefont {Benito~A.}\ \bibnamefont {Ju\'arez-Aubry}}\ and\ \bibinfo {author} {\bibfnamefont {Dimitris}\ \bibnamefont {Moustos}},\ }\bibfield  {title} {\enquote {\bibinfo {title} {Asymptotic states for stationary unruh-dewitt detectors},}\ }\href {\doibase 10.1103/PhysRevD.100.025018} {\bibfield  {journal} {\bibinfo  {journal} {Phys. Rev. D}\ }\textbf {\bibinfo {volume} {100}},\ \bibinfo {pages} {025018} (\bibinfo {year} {2019})}\BibitemShut {NoStop}%
\bibitem [{\citenamefont {{Arias}}\ \emph {et~al.}(2018)\citenamefont {{Arias}}, \citenamefont {{de Oliveira}},\ and\ \citenamefont {{Sarandy}}}]{Arias_2018}%
  \BibitemOpen
  \bibfield  {author} {\bibinfo {author} {\bibfnamefont {Enrique}\ \bibnamefont {{Arias}}}, \bibinfo {author} {\bibfnamefont {Thiago~R.}\ \bibnamefont {{de Oliveira}}}, \ and\ \bibinfo {author} {\bibfnamefont {M.~S.}\ \bibnamefont {{Sarandy}}},\ }\bibfield  {title} {\enquote {\bibinfo {title} {{The Unruh quantum Otto engine}},}\ }\href {\doibase 10.1007/JHEP02(2018)168} {\bibfield  {journal} {\bibinfo  {journal} {Journal of High Energy Physics}\ }\textbf {\bibinfo {volume} {2018}},\ \bibinfo {eid} {168} (\bibinfo {year} {2018})}\BibitemShut {NoStop}%
\bibitem [{\citenamefont {{Gray}}\ and\ \citenamefont {{Mann}}(2018)}]{Mann2018}%
  \BibitemOpen
  \bibfield  {author} {\bibinfo {author} {\bibfnamefont {Finnian}\ \bibnamefont {{Gray}}}\ and\ \bibinfo {author} {\bibfnamefont {Robert~B.}\ \bibnamefont {{Mann}}},\ }\bibfield  {title} {\enquote {\bibinfo {title} {{Scalar and fermionic Unruh Otto engines}},}\ }\href {\doibase 10.1007/JHEP11(2018)174} {\bibfield  {journal} {\bibinfo  {journal} {Journal of High Energy Physics}\ }\textbf {\bibinfo {volume} {2018}},\ \bibinfo {eid} {174} (\bibinfo {year} {2018})}\BibitemShut {NoStop}%
\bibitem [{\citenamefont {{Xu}}\ and\ \citenamefont {{Yung}}(2020)}]{Unruh:Otto:degen}%
  \BibitemOpen
  \bibfield  {author} {\bibinfo {author} {\bibfnamefont {Hao}\ \bibnamefont {{Xu}}}\ and\ \bibinfo {author} {\bibfnamefont {Man-Hong}\ \bibnamefont {{Yung}}},\ }\bibfield  {title} {\enquote {\bibinfo {title} {{Unruh quantum Otto heat engine with level degeneracy}},}\ }\href {\doibase 10.1016/j.physletb.2020.135201} {\bibfield  {journal} {\bibinfo  {journal} {Physics Letters B}\ }\textbf {\bibinfo {volume} {801}},\ \bibinfo {eid} {135201} (\bibinfo {year} {2020})}\BibitemShut {NoStop}%
\bibitem [{\citenamefont {{Kane}}\ and\ \citenamefont {{Majhi}}(2021)}]{Unruh:Otto:entangl}%
  \BibitemOpen
  \bibfield  {author} {\bibinfo {author} {\bibfnamefont {Gaurang~Ramakant}\ \bibnamefont {{Kane}}}\ and\ \bibinfo {author} {\bibfnamefont {Bibhas~Ranjan}\ \bibnamefont {{Majhi}}},\ }\bibfield  {title} {\enquote {\bibinfo {title} {{Entangled quantum Unruh Otto engine is more efficient}},}\ }\href {\doibase 10.1103/PhysRevD.104.L041701} {\bibfield  {journal} {\bibinfo  {journal} {\prd}\ }\textbf {\bibinfo {volume} {104}},\ \bibinfo {eid} {L041701} (\bibinfo {year} {2021})}\BibitemShut {NoStop}%
\bibitem [{\citenamefont {{Barman}}\ and\ \citenamefont {{Majhi}}(2022)}]{Unruh:Otto:entangl2}%
  \BibitemOpen
  \bibfield  {author} {\bibinfo {author} {\bibfnamefont {Dipankar}\ \bibnamefont {{Barman}}}\ and\ \bibinfo {author} {\bibfnamefont {Bibhas~Ranjan}\ \bibnamefont {{Majhi}}},\ }\bibfield  {title} {\enquote {\bibinfo {title} {{Constructing an entangled Unruh Otto engine and its efficiency}},}\ }\href {\doibase 10.1007/JHEP05(2022)046} {\bibfield  {journal} {\bibinfo  {journal} {Journal of High Energy Physics}\ }\textbf {\bibinfo {volume} {2022}},\ \bibinfo {eid} {46} (\bibinfo {year} {2022})}\BibitemShut {NoStop}%
\bibitem [{\citenamefont {{Mukherjee}}\ \emph {et~al.}(2022)\citenamefont {{Mukherjee}}, \citenamefont {{Gangopadhyay}},\ and\ \citenamefont {{Majumdar}}}]{Unruh:Otto:reflect}%
  \BibitemOpen
  \bibfield  {author} {\bibinfo {author} {\bibfnamefont {Arnab}\ \bibnamefont {{Mukherjee}}}, \bibinfo {author} {\bibfnamefont {Sunandan}\ \bibnamefont {{Gangopadhyay}}}, \ and\ \bibinfo {author} {\bibfnamefont {A.~S.}\ \bibnamefont {{Majumdar}}},\ }\bibfield  {title} {\enquote {\bibinfo {title} {{Unruh quantum Otto engine in the presence of a reflecting boundary}},}\ }\href {\doibase 10.1007/JHEP09(2022)105} {\bibfield  {journal} {\bibinfo  {journal} {Journal of High Energy Physics}\ }\textbf {\bibinfo {volume} {2022}},\ \bibinfo {eid} {105} (\bibinfo {year} {2022})}\BibitemShut {NoStop}%
\bibitem [{\citenamefont {{Myers}}\ \emph {et~al.}(2021)\citenamefont {{Myers}}, \citenamefont {{Abah}},\ and\ \citenamefont {{Deffner}}}]{Otto:relativ:energ}%
  \BibitemOpen
  \bibfield  {author} {\bibinfo {author} {\bibfnamefont {Nathan~M.}\ \bibnamefont {{Myers}}}, \bibinfo {author} {\bibfnamefont {Obinna}\ \bibnamefont {{Abah}}}, \ and\ \bibinfo {author} {\bibfnamefont {Sebastian}\ \bibnamefont {{Deffner}}},\ }\bibfield  {title} {\enquote {\bibinfo {title} {{Quantum Otto engines at relativistic energies}},}\ }\href {\doibase 10.1088/1367-2630/ac2756} {\bibfield  {journal} {\bibinfo  {journal} {New Journal of Physics}\ }\textbf {\bibinfo {volume} {23}},\ \bibinfo {eid} {105001} (\bibinfo {year} {2021})}\BibitemShut {NoStop}%
\bibitem [{\citenamefont {{Ferketic}}\ and\ \citenamefont {{Deffner}}(2023)}]{Deffner:spacetime}%
  \BibitemOpen
  \bibfield  {author} {\bibinfo {author} {\bibfnamefont {Emily~E.}\ \bibnamefont {{Ferketic}}}\ and\ \bibinfo {author} {\bibfnamefont {Sebastian}\ \bibnamefont {{Deffner}}},\ }\bibfield  {title} {\enquote {\bibinfo {title} {{Boosting thermodynamic performance by bending space-time}},}\ }\href {\doibase 10.1209/0295-5075/acad9c} {\bibfield  {journal} {\bibinfo  {journal} {EPL (Europhysics Letters)}\ }\textbf {\bibinfo {volume} {141}},\ \bibinfo {eid} {19001} (\bibinfo {year} {2023})}\BibitemShut {NoStop}%
\bibitem [{\citenamefont {{Bruschi}}\ \emph {et~al.}(2020)\citenamefont {{Bruschi}}, \citenamefont {{Morris}},\ and\ \citenamefont {{Fuentes}}}]{DEB1}%
  \BibitemOpen
  \bibfield  {author} {\bibinfo {author} {\bibfnamefont {David~Edward}\ \bibnamefont {{Bruschi}}}, \bibinfo {author} {\bibfnamefont {Benjamin}\ \bibnamefont {{Morris}}}, \ and\ \bibinfo {author} {\bibfnamefont {Ivette}\ \bibnamefont {{Fuentes}}},\ }\bibfield  {title} {\enquote {\bibinfo {title} {{Thermodynamics of relativistic quantum fields confined in cavities}},}\ }\href {\doibase 10.1016/j.physleta.2020.126601} {\bibfield  {journal} {\bibinfo  {journal} {Physics Letters A}\ }\textbf {\bibinfo {volume} {384}},\ \bibinfo {eid} {126601} (\bibinfo {year} {2020})}\BibitemShut {NoStop}%
\bibitem [{\citenamefont {Ferreri}\ \emph {et~al.}(2023)\citenamefont {Ferreri}, \citenamefont {Macr\`{\i}}, \citenamefont {Wilhelm}, \citenamefont {Nori},\ and\ \citenamefont {Bruschi}}]{DEB2}%
  \BibitemOpen
  \bibfield  {author} {\bibinfo {author} {\bibfnamefont {Alessandro}\ \bibnamefont {Ferreri}}, \bibinfo {author} {\bibfnamefont {Vincenzo}\ \bibnamefont {Macr\`{\i}}}, \bibinfo {author} {\bibfnamefont {Frank~K.}\ \bibnamefont {Wilhelm}}, \bibinfo {author} {\bibfnamefont {Franco}\ \bibnamefont {Nori}}, \ and\ \bibinfo {author} {\bibfnamefont {David~Edward}\ \bibnamefont {Bruschi}},\ }\bibfield  {title} {\enquote {\bibinfo {title} {Quantum field heat engine powered by phonon-photon interactions},}\ }\href {\doibase 10.1103/PhysRevResearch.5.043274} {\bibfield  {journal} {\bibinfo  {journal} {Phys. Rev. Res.}\ }\textbf {\bibinfo {volume} {5}},\ \bibinfo {pages} {043274} (\bibinfo {year} {2023})}\BibitemShut {NoStop}%
\bibitem [{\citenamefont {{Papadatos}}(2021)}]{papadatos}%
  \BibitemOpen
  \bibfield  {author} {\bibinfo {author} {\bibfnamefont {Nikolaos}\ \bibnamefont {{Papadatos}}},\ }\bibfield  {title} {\enquote {\bibinfo {title} {{The Quantum Otto Heat Engine with a Relativistically Moving Thermal Bath}},}\ }\href {\doibase 10.1007/s10773-021-04969-9} {\bibfield  {journal} {\bibinfo  {journal} {International Journal of Theoretical Physics}\ }\textbf {\bibinfo {volume} {60}},\ \bibinfo {pages} {4210--4223} (\bibinfo {year} {2021})}\BibitemShut {NoStop}%
\bibitem [{\citenamefont {{Good}}\ \emph {et~al.}(2020)\citenamefont {{Good}}, \citenamefont {{Ju{\'a}rez-Aubry}}, \citenamefont {{Moustos}},\ and\ \citenamefont {{Temirkhan}}}]{Effective:Unruh}%
  \BibitemOpen
  \bibfield  {author} {\bibinfo {author} {\bibfnamefont {Michael}\ \bibnamefont {{Good}}}, \bibinfo {author} {\bibfnamefont {Benito~A.}\ \bibnamefont {{Ju{\'a}rez-Aubry}}}, \bibinfo {author} {\bibfnamefont {Dimitris}\ \bibnamefont {{Moustos}}}, \ and\ \bibinfo {author} {\bibfnamefont {Maksat}\ \bibnamefont {{Temirkhan}}},\ }\bibfield  {title} {\enquote {\bibinfo {title} {{Unruh-like effects: effective temperatures along stationary worldlines}},}\ }\href {\doibase 10.1007/JHEP06(2020)059} {\bibfield  {journal} {\bibinfo  {journal} {Journal of High Energy Physics}\ }\textbf {\bibinfo {volume} {2020}},\ \bibinfo {eid} {59} (\bibinfo {year} {2020})}\BibitemShut {NoStop}%
\bibitem [{\citenamefont {{Gallock-Yoshimura}}\ \emph {et~al.}(2023)\citenamefont {{Gallock-Yoshimura}}, \citenamefont {{Thakur}},\ and\ \citenamefont {{Mann}}}]{Yoshimura:mann}%
  \BibitemOpen
  \bibfield  {author} {\bibinfo {author} {\bibfnamefont {Kensuke}\ \bibnamefont {{Gallock-Yoshimura}}}, \bibinfo {author} {\bibfnamefont {Vaishant}\ \bibnamefont {{Thakur}}}, \ and\ \bibinfo {author} {\bibfnamefont {Robert~B.}\ \bibnamefont {{Mann}}},\ }\bibfield  {title} {\enquote {\bibinfo {title} {{Quantum Otto engine driven by quantum fields}},}\ }\href {\doibase 10.3389/fphy.2023.1287860} {\bibfield  {journal} {\bibinfo  {journal} {Frontiers in Physics}\ }\textbf {\bibinfo {volume} {11}},\ \bibinfo {eid} {1287860} (\bibinfo {year} {2023})}\BibitemShut {NoStop}%
\bibitem [{\citenamefont {DeWitt}(1979)}]{DeWitt}%
  \BibitemOpen
  \bibfield  {author} {\bibinfo {author} {\bibfnamefont {B.~S.}\ \bibnamefont {DeWitt}},\ }\bibfield  {title} {\enquote {\bibinfo {title} {Quantum gravity: The new synthesis},}\ }in\ \href@noop {} {\emph {\bibinfo {booktitle} {General Relativity: an Einstein Centenary Survey}}},\ \bibinfo {editor} {edited by\ \bibinfo {editor} {\bibfnamefont {S.}~\bibnamefont {Hawking}}\ and\ \bibinfo {editor} {\bibfnamefont {W.}~\bibnamefont {Israel}}}\ (\bibinfo  {publisher} {Cambridge University Press},\ \bibinfo {address} {Cambridge, England},\ \bibinfo {year} {1979})\BibitemShut {NoStop}%
\bibitem [{\citenamefont {{Hu}}\ \emph {et~al.}(2012)\citenamefont {{Hu}}, \citenamefont {{Lin}},\ and\ \citenamefont {{Louko}}}]{Hu:Louko}%
  \BibitemOpen
  \bibfield  {author} {\bibinfo {author} {\bibfnamefont {B.~L.}\ \bibnamefont {{Hu}}}, \bibinfo {author} {\bibfnamefont {Shih-Yuin}\ \bibnamefont {{Lin}}}, \ and\ \bibinfo {author} {\bibfnamefont {Jorma}\ \bibnamefont {{Louko}}},\ }\bibfield  {title} {\enquote {\bibinfo {title} {{Relativistic quantum information in detectors-field interactions}},}\ }\href {https://iopscience.iop.org/article/10.1088/0264-9381/29/22/224005} {\bibfield  {journal} {\bibinfo  {journal} {Classical and Quantum Gravity}\ }\textbf {\bibinfo {volume} {29}},\ \bibinfo {eid} {224005} (\bibinfo {year} {2012})}\BibitemShut {NoStop}%
\bibitem [{\citenamefont {{Simidzija}}\ and\ \citenamefont {{Mart{\'\i}n-Mart{\'\i}nez}}(2017)}]{delta1}%
  \BibitemOpen
  \bibfield  {author} {\bibinfo {author} {\bibfnamefont {Petar}\ \bibnamefont {{Simidzija}}}\ and\ \bibinfo {author} {\bibfnamefont {Eduardo}\ \bibnamefont {{Mart{\'\i}n-Mart{\'\i}nez}}},\ }\bibfield  {title} {\enquote {\bibinfo {title} {{Nonperturbative analysis of entanglement harvesting from coherent field states}},}\ }\href {\doibase 10.1103/PhysRevD.96.065008} {\bibfield  {journal} {\bibinfo  {journal} {\prd}\ }\textbf {\bibinfo {volume} {96}},\ \bibinfo {eid} {065008} (\bibinfo {year} {2017})}\BibitemShut {NoStop}%
\bibitem [{\citenamefont {{de Ram{\'o}n}}\ and\ \citenamefont {{Martin-Martinez}}(2020)}]{delta2}%
  \BibitemOpen
  \bibfield  {author} {\bibinfo {author} {\bibfnamefont {Jos{\'e}}\ \bibnamefont {{de Ram{\'o}n}}}\ and\ \bibinfo {author} {\bibfnamefont {Eduardo}\ \bibnamefont {{Martin-Martinez}}},\ }\bibfield  {title} {\enquote {\bibinfo {title} {{A non-perturbative analysis of spin-boson interactions using the Weyl relations}},}\ }\href {\doibase 10.48550/arXiv.2002.01994} {\bibfield  {journal} {\bibinfo  {journal} {arXiv e-prints}\ ,\ \bibinfo {eid} {arXiv:2002.01994}} (\bibinfo {year} {2020})},\ \Eprint {http://arxiv.org/abs/2002.01994} {arXiv:2002.01994 [quant-ph]} \BibitemShut {NoStop}%
\bibitem [{\citenamefont {Simidzija}\ \emph {et~al.}(2020)\citenamefont {Simidzija}, \citenamefont {Ahmadzadegan}, \citenamefont {Kempf},\ and\ \citenamefont {Mart\'{\i}n-Mart\'{\i}nez}}]{delta3}%
  \BibitemOpen
  \bibfield  {author} {\bibinfo {author} {\bibfnamefont {Petar}\ \bibnamefont {Simidzija}}, \bibinfo {author} {\bibfnamefont {Aida}\ \bibnamefont {Ahmadzadegan}}, \bibinfo {author} {\bibfnamefont {Achim}\ \bibnamefont {Kempf}}, \ and\ \bibinfo {author} {\bibfnamefont {Eduardo}\ \bibnamefont {Mart\'{\i}n-Mart\'{\i}nez}},\ }\bibfield  {title} {\enquote {\bibinfo {title} {Transmission of quantum information through quantum fields},}\ }\href {\doibase 10.1103/PhysRevD.101.036014} {\bibfield  {journal} {\bibinfo  {journal} {Phys. Rev. D}\ }\textbf {\bibinfo {volume} {101}},\ \bibinfo {pages} {036014} (\bibinfo {year} {2020})}\BibitemShut {NoStop}%
\bibitem [{\citenamefont {Gallock-Yoshimura}\ and\ \citenamefont {Mann}(2021)}]{delta4}%
  \BibitemOpen
  \bibfield  {author} {\bibinfo {author} {\bibfnamefont {Kensuke}\ \bibnamefont {Gallock-Yoshimura}}\ and\ \bibinfo {author} {\bibfnamefont {Robert~B.}\ \bibnamefont {Mann}},\ }\bibfield  {title} {\enquote {\bibinfo {title} {Entangled detectors nonperturbatively harvest mutual information},}\ }\href {\doibase 10.1103/PhysRevD.104.125017} {\bibfield  {journal} {\bibinfo  {journal} {Phys. Rev. D}\ }\textbf {\bibinfo {volume} {104}},\ \bibinfo {pages} {125017} (\bibinfo {year} {2021})}\BibitemShut {NoStop}%
\bibitem [{\citenamefont {Tjoa}\ and\ \citenamefont {Gallock-Yoshimura}(2022)}]{delta5}%
  \BibitemOpen
  \bibfield  {author} {\bibinfo {author} {\bibfnamefont {Erickson}\ \bibnamefont {Tjoa}}\ and\ \bibinfo {author} {\bibfnamefont {Kensuke}\ \bibnamefont {Gallock-Yoshimura}},\ }\bibfield  {title} {\enquote {\bibinfo {title} {Channel capacity of relativistic quantum communication with rapid interaction},}\ }\href {\doibase 10.1103/PhysRevD.105.085011} {\bibfield  {journal} {\bibinfo  {journal} {Phys. Rev. D}\ }\textbf {\bibinfo {volume} {105}},\ \bibinfo {pages} {085011} (\bibinfo {year} {2022})}\BibitemShut {NoStop}%
\bibitem [{\citenamefont {{M{\'e}ndez Avalos}}\ \emph {et~al.}(2022)\citenamefont {{M{\'e}ndez Avalos}}, \citenamefont {{Gallock-Yoshimura}}, \citenamefont {{Henderson}},\ and\ \citenamefont {{Mann}}}]{delta6}%
  \BibitemOpen
  \bibfield  {author} {\bibinfo {author} {\bibfnamefont {Diana}\ \bibnamefont {{M{\'e}ndez Avalos}}}, \bibinfo {author} {\bibfnamefont {Kensuke}\ \bibnamefont {{Gallock-Yoshimura}}}, \bibinfo {author} {\bibfnamefont {Laura~J.}\ \bibnamefont {{Henderson}}}, \ and\ \bibinfo {author} {\bibfnamefont {Robert~B.}\ \bibnamefont {{Mann}}},\ }\bibfield  {title} {\enquote {\bibinfo {title} {{Instant Extraction of Non-Perturbative Tripartite Entanglement}},}\ }\href {\doibase 10.48550/arXiv.2204.02983} {\bibfield  {journal} {\bibinfo  {journal} {arXiv e-prints}\ ,\ \bibinfo {eid} {arXiv:2204.02983}} (\bibinfo {year} {2022})},\ \Eprint {http://arxiv.org/abs/2204.02983} {arXiv:2204.02983 [quant-ph]} \BibitemShut {NoStop}%
\bibitem [{\citenamefont {Kollas}\ \emph {et~al.}(2023)\citenamefont {Kollas}, \citenamefont {Moustos},\ and\ \citenamefont {Mu\~noz}}]{NK:DM:MM}%
  \BibitemOpen
  \bibfield  {author} {\bibinfo {author} {\bibfnamefont {Nikolaos~K.}\ \bibnamefont {Kollas}}, \bibinfo {author} {\bibfnamefont {Dimitris}\ \bibnamefont {Moustos}}, \ and\ \bibinfo {author} {\bibfnamefont {Miguel~R.}\ \bibnamefont {Mu\~noz}},\ }\bibfield  {title} {\enquote {\bibinfo {title} {Cohering and decohering power of massive scalar fields under instantaneous interactions},}\ }\href {\doibase 10.1103/PhysRevA.107.022420} {\bibfield  {journal} {\bibinfo  {journal} {Phys. Rev. A}\ }\textbf {\bibinfo {volume} {107}},\ \bibinfo {pages} {022420} (\bibinfo {year} {2023})}\BibitemShut {NoStop}%
\bibitem [{\citenamefont {Horodecki}\ \emph {et~al.}(2003)\citenamefont {Horodecki}, \citenamefont {Horodecki},\ and\ \citenamefont {Oppenheim}}]{PhysRevA.67.062104}%
  \BibitemOpen
  \bibfield  {author} {\bibinfo {author} {\bibfnamefont {Micha\l{}}\ \bibnamefont {Horodecki}}, \bibinfo {author} {\bibfnamefont {Pawe\l{}}\ \bibnamefont {Horodecki}}, \ and\ \bibinfo {author} {\bibfnamefont {Jonathan}\ \bibnamefont {Oppenheim}},\ }\bibfield  {title} {\enquote {\bibinfo {title} {Reversible transformations from pure to mixed states and the unique measure of information},}\ }\href {\doibase 10.1103/PhysRevA.67.062104} {\bibfield  {journal} {\bibinfo  {journal} {Phys. Rev. A}\ }\textbf {\bibinfo {volume} {67}},\ \bibinfo {pages} {062104} (\bibinfo {year} {2003})}\BibitemShut {NoStop}%
\bibitem [{\citenamefont {Schlicht}(2004)}]{Schlicht_2004}%
  \BibitemOpen
  \bibfield  {author} {\bibinfo {author} {\bibfnamefont {Sebastian}\ \bibnamefont {Schlicht}},\ }\bibfield  {title} {\enquote {\bibinfo {title} {Considerations on the unruh effect: causality and regularization},}\ }\href {\doibase 10.1088/0264-9381/21/19/011} {\bibfield  {journal} {\bibinfo  {journal} {Classical and Quantum Gravity}\ }\textbf {\bibinfo {volume} {21}},\ \bibinfo {pages} {4647--4660} (\bibinfo {year} {2004})}\BibitemShut {NoStop}%
\bibitem [{\citenamefont {Louko}\ and\ \citenamefont {Satz}(2006)}]{Louko_2006}%
  \BibitemOpen
  \bibfield  {author} {\bibinfo {author} {\bibfnamefont {Jorma}\ \bibnamefont {Louko}}\ and\ \bibinfo {author} {\bibfnamefont {Alejandro}\ \bibnamefont {Satz}},\ }\bibfield  {title} {\enquote {\bibinfo {title} {How often does the unruh{\textendash}{DeWitt} detector click? regularization by a spatial profile},}\ }\href {\doibase 10.1088/0264-9381/23/22/015} {\bibfield  {journal} {\bibinfo  {journal} {Classical and Quantum Gravity}\ }\textbf {\bibinfo {volume} {23}},\ \bibinfo {pages} {6321--6343} (\bibinfo {year} {2006})}\BibitemShut {NoStop}%
\bibitem [{\citenamefont {Pozas-Kerstjens}\ and\ \citenamefont {Mart\'{\i}n-Mart\'{\i}nez}(2016)}]{smear:mart1}%
  \BibitemOpen
  \bibfield  {author} {\bibinfo {author} {\bibfnamefont {Alejandro}\ \bibnamefont {Pozas-Kerstjens}}\ and\ \bibinfo {author} {\bibfnamefont {Eduardo}\ \bibnamefont {Mart\'{\i}n-Mart\'{\i}nez}},\ }\bibfield  {title} {\enquote {\bibinfo {title} {Entanglement harvesting from the electromagnetic vacuum with hydrogenlike atoms},}\ }\href {\doibase 10.1103/PhysRevD.94.064074} {\bibfield  {journal} {\bibinfo  {journal} {Phys. Rev. D}\ }\textbf {\bibinfo {volume} {94}},\ \bibinfo {pages} {064074} (\bibinfo {year} {2016})}\BibitemShut {NoStop}%
\bibitem [{\citenamefont {Perche}(2022)}]{T:Rick}%
  \BibitemOpen
  \bibfield  {author} {\bibinfo {author} {\bibfnamefont {T.~Rick}\ \bibnamefont {Perche}},\ }\bibfield  {title} {\enquote {\bibinfo {title} {Localized nonrelativistic quantum systems in curved spacetimes: A general characterization of particle detector models},}\ }\href {\doibase 10.1103/PhysRevD.106.025018} {\bibfield  {journal} {\bibinfo  {journal} {Phys. Rev. D}\ }\textbf {\bibinfo {volume} {106}},\ \bibinfo {pages} {025018} (\bibinfo {year} {2022})}\BibitemShut {NoStop}%
\bibitem [{\citenamefont {Simidzija}\ \emph {et~al.}(2018)\citenamefont {Simidzija}, \citenamefont {Jonsson},\ and\ \citenamefont {Mart\'{\i}n-Mart\'{\i}nez}}]{Chlimitz:nogo}%
  \BibitemOpen
  \bibfield  {author} {\bibinfo {author} {\bibfnamefont {Petar}\ \bibnamefont {Simidzija}}, \bibinfo {author} {\bibfnamefont {Robert~H.}\ \bibnamefont {Jonsson}}, \ and\ \bibinfo {author} {\bibfnamefont {Eduardo}\ \bibnamefont {Mart\'{\i}n-Mart\'{\i}nez}},\ }\bibfield  {title} {\enquote {\bibinfo {title} {General no-go theorem for entanglement extraction},}\ }\href {\doibase 10.1103/PhysRevD.97.125002} {\bibfield  {journal} {\bibinfo  {journal} {Phys. Rev. D}\ }\textbf {\bibinfo {volume} {97}},\ \bibinfo {pages} {125002} (\bibinfo {year} {2018})}\BibitemShut {NoStop}%
\bibitem [{\citenamefont {Kollas}\ and\ \citenamefont {Moustos}(2022)}]{PhysRevD.105.025006}%
  \BibitemOpen
  \bibfield  {author} {\bibinfo {author} {\bibfnamefont {Nikolaos~K.}\ \bibnamefont {Kollas}}\ and\ \bibinfo {author} {\bibfnamefont {Dimitris}\ \bibnamefont {Moustos}},\ }\bibfield  {title} {\enquote {\bibinfo {title} {Generation and catalysis of coherence with scalar fields},}\ }\href {\doibase 10.1103/PhysRevD.105.025006} {\bibfield  {journal} {\bibinfo  {journal} {Phys. Rev. D}\ }\textbf {\bibinfo {volume} {105}},\ \bibinfo {pages} {025006} (\bibinfo {year} {2022})}\BibitemShut {NoStop}%
\bibitem [{\citenamefont {Onoe}\ \emph {et~al.}(2022)\citenamefont {Onoe}, \citenamefont {Guedes}, \citenamefont {Moskalenko}, \citenamefont {Leitenstorfer}, \citenamefont {Burkard},\ and\ \citenamefont {Ralph}}]{PhysRevD.105.056023}%
  \BibitemOpen
  \bibfield  {author} {\bibinfo {author} {\bibfnamefont {Sho}\ \bibnamefont {Onoe}}, \bibinfo {author} {\bibfnamefont {Thiago L.~M.}\ \bibnamefont {Guedes}}, \bibinfo {author} {\bibfnamefont {Andrey~S.}\ \bibnamefont {Moskalenko}}, \bibinfo {author} {\bibfnamefont {Alfred}\ \bibnamefont {Leitenstorfer}}, \bibinfo {author} {\bibfnamefont {Guido}\ \bibnamefont {Burkard}}, \ and\ \bibinfo {author} {\bibfnamefont {Timothy~C.}\ \bibnamefont {Ralph}},\ }\bibfield  {title} {\enquote {\bibinfo {title} {Realizing a rapidly switched unruh-dewitt detector through electro-optic sampling of the electromagnetic vacuum},}\ }\href {\doibase 10.1103/PhysRevD.105.056023} {\bibfield  {journal} {\bibinfo  {journal} {Phys. Rev. D}\ }\textbf {\bibinfo {volume} {105}},\ \bibinfo {pages} {056023} (\bibinfo {year} {2022})}\BibitemShut {NoStop}%
\bibitem [{\citenamefont {Gallock-Yoshimura}(2024)}]{gallockyoshimura2023relativistic}%
  \BibitemOpen
  \bibfield  {author} {\bibinfo {author} {\bibfnamefont {Kensuke}\ \bibnamefont {Gallock-Yoshimura}},\ }\bibfield  {title} {\enquote {\bibinfo {title} {Relativistic quantum otto engine: instant work extraction from a quantum field},}\ }\href {https://link.springer.com/article/10.1007/JHEP01(2024)198} {\bibfield  {journal} {\bibinfo  {journal} {Journal of High Energy Physics}\ }\textbf {\bibinfo {volume} {2024}},\ \bibinfo {pages} {198} (\bibinfo {year} {2024})}\BibitemShut {NoStop}%
\bibitem [{\citenamefont {Polo-G\'omez}\ and\ \citenamefont {Mart\'{\i}n-Mart\'{\i}nez}(2024)}]{pologomez2023nonperturbative}%
  \BibitemOpen
  \bibfield  {author} {\bibinfo {author} {\bibfnamefont {Jos\'e}\ \bibnamefont {Polo-G\'omez}}\ and\ \bibinfo {author} {\bibfnamefont {Eduardo}\ \bibnamefont {Mart\'{\i}n-Mart\'{\i}nez}},\ }\bibfield  {title} {\enquote {\bibinfo {title} {Nonperturbative method for particle detectors with continuous interactions},}\ }\href {\doibase 10.1103/PhysRevD.109.045014} {\bibfield  {journal} {\bibinfo  {journal} {Phys. Rev. D}\ }\textbf {\bibinfo {volume} {109}},\ \bibinfo {pages} {045014} (\bibinfo {year} {2024})}\BibitemShut {NoStop}%
\end{thebibliography}%
\end{document}